\documentclass[a4paper,11pt]{article}
\pdfoutput=1
\usepackage{jheppub}

\usepackage{amsmath}
\usepackage{mathtools}
\usepackage{amssymb}
\usepackage{amsthm}
\usepackage{physics}
\usepackage{braket}
\usepackage{tikz-cd}
\usepackage{nicematrix}

\usepackage[normalem]{ulem}

\numberwithin{equation}{section}

\renewcommand{\thefigure}{\arabic{section}.\arabic{figure}}

\DeclareMathOperator{\Span}{Span}

\usetikzlibrary{decorations.pathreplacing,decorations.markings}
\tikzset{
    set arrow inside/.code={\pgfqkeys{/tikz/arrow inside}{#1}},
    set arrow inside={end/.initial=>, opt/.initial=},
    /pgf/decoration/Mark/.style={
        mark/.expanded=at position #1 with
        {
            \noexpand\arrow[\pgfkeysvalueof{/tikz/arrow inside/opt}]{\pgfkeysvalueof{/tikz/arrow inside/end}}
        }
    },
    arrow inside/.style 2 args={
        set arrow inside={#1},
        postaction={
            decorate,decoration={
                markings,Mark/.list={#2}
            }
        }
    },
}

\newcommand*\diff{\mathop{}\!\mathrm{d}}

\title{\LARGE 
An Interacting, Higher Derivative, Boundary Conformal Field Theory
}

\author[a]{Christopher P.\ Herzog,}
\author[b]{Yanjun Zhou}
\affiliation[a]{Department of Mathematics, King's College London, Strand, London, WC2R 2LS, UK}
\affiliation[b]{Rudolf Peierls Centre for Theoretical Physics, University of Oxford, Parks Road, Oxford, OX1 3PU, UK}

\emailAdd{christopher.herzog@kcl.ac.uk}
\emailAdd{yanjunzhou2001@gmail.com}

\abstract{
We consider a higher derivative scalar field theory in the presence of a boundary and a classically marginal interaction.  We first investigate the free limit where the scalar obeys the square of the Klein-Gordon equation.  In precisely $d=6$ dimensions, modules generated by $d-2$ and $d-4$ dimensional primaries merge to form a staggered module.  We compute the conformal block associated with this module and show that it is a generalized eigenvector of the Casimir operator.  
Next we include the effect of a classically marginal interaction that involves four scalar fields and two derivatives.  The theory has an infrared fixed point in $d=6-\epsilon$ dimensions.  We compute boundary operator anomalous dimensions and boundary OPE coefficients at leading order in the $\epsilon$ expansion for the allowed conformal boundary conditions.

}

\makeatletter
\def\@fpheader{\vspace{0cm}}
\makeatother

\begin{document}
\maketitle

\section{Introduction}

We deform a higher derivative free scalar field theory by including a boundary and a classically marginal interaction.
Theories with any two of these features have been studied before in the literature.  
An elaborate set of conformal symmetry preserving boundary conditions were worked out in ref.\ \cite{Chalabi:2022qit} for higher derivative free scalar field theories.  Anomalous dimensions in the epsilon expansion 
for interacting higher derivative scalar field theories are in \cite{dengler1985renormalization,aharony1985novel,aharony1987renormalization,Gliozzi:2016ysv, Gliozzi:2017hni, Safari:2017irw, Safari:2017tgs,Safari:2021ocb}. 
Studying interacting scalar field theories in the presence of a boundary and a standard $\phi \Box \phi$ kinetic term has a long and illustrious history.  See ref.\ \cite{Diehl:1996kd} for a recent review.  
The closest work to this one is perhaps ref.\ \cite{Diehl:2006klj} where the authors consider Lifshitz theories with a boundary, including fourth order derivative kinetic terms in some but not all directions, and adding a different type of interaction term than the one we consider.  

To be concrete, we investigate the action
\begin{equation}
\label{action}
S = \int_{x_n\geq0} \diff^d x \left[ \phi \Box^2 \phi + g \mu^{d-\Delta} \left(
\phi^3 \Box \phi + \frac{2}{d-4} \phi^2 (\partial \phi)^2 
\right)
\right] \ .
\end{equation}
Here $\phi$ is the scalar field, $x_n$ is the direction normal to the boundary,
$\mu$ is an energy scale, $g$ is the interaction strength, and $\Delta$ is the scaling dimension of the quartic operator, classically equal to $2d-6$. 
The quartic interaction is a primary operator in the theory with $g=0$, and we will study the response of the system to its presence in the epsilon expansion, with $d = 6-\epsilon$.

What can be gained by looking at all three conditions at once -- higher derivatives, boundary, and an interaction -- that cannot be extracted by considering these deformations two at a time, or indeed even one at a time?
Cosmology may be the most natural framework in which all three elements interact.  
For example Starobinsky famously argued that in the early universe quantum effects should lead to higher derivative corrections to Einstein's equations.  His $R + R^2$ theory of gravity naturally produces an inflationary de Sitter epoch in the history of the universe \cite{Starobinsky:1979ty}.\footnote{
One may reformulate Starobinsky inflation using an auxiliary scalar field and a two derivative action, but the point remains that quantum effects generically lead to higher derivative terms.
}
Although tackling theories of quantum gravity and even quantum field theory are difficult in de Sitter space, a target of opportunity is the boundary at future infinity. Using this boundary, for example, one \cite{Strominger:2001pn} might hope to generalize the AdS/CFT correspondence; indeed most formulations of dS/CFT involve non-unitary conformal field theory not unlike the one considered here. 

Our main motivation for considering (\ref{action}) is thus conceptual.  It is a simple, tractable example of a non-unitary conformal field theory (CFT) that may shed light on issues in cosmology.  The model is non-unitary because the scaling dimension of the scalar field $\phi$ sits well below the unitarity bound $\frac{d-2}{2}$ in perturbation theory.

A more pedestrian motivation is condensed matter experiment.  There are materials that go through both bulk and surface ordering phase transitions where associated critical exponents can be measured.  If the spatial part of the kinetic term is tuned to zero, higher derivative contributions can become dominant giving rise to so-called Lifshitz theories
\cite{diehl2002critical}.  
Perhaps more familiar in an engineering context, elastic models of plate bending are also often formulated in a higher derivative framework \cite{landau1986theory}.  While our $d=6-\epsilon$ example is probably well away from any experimental regime, unless somehow $\epsilon$ can be set to two or three at the end of the perturbative calculation, 
some of the lessons we learn here may be useful in analyzing models with $2 \leq d \leq 4$
and thus be relevant for experiment.

To say a little more in this condensed matter context, our action (\ref{action}) is a special case of a Lifshiftz tricritical theory.  If $m$ directions are associated with fourth order derivative kinetic terms and $d-m$ directions with second order derivative kinetic terms, our case sets $m=d=6-\epsilon$, but more generally the upper critical dimension is reduced from 6 to $d = 3+m/2$. 
In this Lifshitz context, the same quartic interaction term in (\ref{action}) was considered already by refs.\ \cite{dengler1985renormalization,aharony1985novel,aharony1987renormalization}.  They had a specific experimental application in mind, RbCaF$_3$.  
The perovskite undergoes a structural antiferrodistortive phase transition.  The usual transition point is very close to a Lifshitz point, and the two can be made coincident through uniaxial pressure \cite{aharony1979lifshitz,buzare1979tricritical,muller1980shift}.  More recently, ref.\ \cite{dutta2013lifshitz} 
has argued for a connection between the Lifshitz tricritical phase transition and a phase transition to an FFLO superconducting state; further, a tricritical Lifshitz point has been observed in Sn$_2$P$_2$(S$_{1-x}$Se$_x$) at $x=0.28$ doped with Ge and Pb \cite{oleaga2017search}.
A natural extension of these earlier works is to consider the effects of boundaries on Lifshitz tricritical systems. (Note ref.\ \cite{Diehl:2006klj} mentioned above focuses on Lifshitz critical rather than Lifshitz tricritical systems.)

We present two results in this work.  The first is a detailed analysis of what happens in the free limit at and slightly below $d=6$.  We are able to extend and flesh out an observation in ref.\ \cite{Brust:2016gjy} about the existence of a staggered  module. The bulk conformal block decomposition of $\braket{\phi(x)\phi(0)}$ becomes singular in the limit
$d \to 6$.  A divergence in the $\phi^2$ conformal block is canceled by a divergence in the OPE coefficient of the dimension four primary, which schematically we can write as $\partial^2 \phi^2$.  Indeed, in precisely $d=6$, $\partial^2 \phi^2$ becomes a descendant of $\phi^2$. This merger of two conformal blocks
leads to constraints on the conformal data for
our interacting model, for example setting the $O(\epsilon)$ corrections to their anomalous dimensions equal.
Our analysis may be useful more widely.  Staggered modules appear in the study of $2d$ logarithmic CFTs
\cite{Rohsiepe:1996qj,Gaberdiel:2001tr,Kytola:2009ax,Creutzig:2013hma}, $2d$ Carrollian CFTs \cite{Hao:2021urq,Hao:2022xhq,Yu:2022bcp,Chen:2023pqf}, and $2d$ celestial CFTs \cite{Chang:2023ttm}. 

The second result is a list of boundary anomalous dimensions and OPE coefficients for the interacting theory (\ref{action}).  
As there are four distinct boundary conditions available for this
$\phi \Box^2 \phi$ theory \cite{Chalabi:2022qit}, we are able to give four distinct sets of perturbative conformal data.  As mentioned above, while it's unlikely the continuation to $d=4$ is meaningful, we hope the techniques we used in these calculations may be more generally employed.  In particular we apply a combination of crossing symmetry and the equations of motion to correlation functions, a strategy 
pioneered in the unitary case with boundary by ref.\ \cite{Giombi:2020rmc} and later applied to fermions in refs.\ \cite{Giombi:2021cnr,Herzog:2022jlx}.

In our interacting model, the $d=6-\epsilon$ bulk physics is more involved than the $d=5-\epsilon$ boundary physics due to the epsilon expansion around the $d=6$ staggered module. Hence, an additional motivation to introduce the boundary is to reconstruct the bulk physics from the boundary physics using crossing symmetry.

Our interaction is the composite operator $\phi^2 \cdot \partial^2\phi^2$, which is the product of the two free-theory primaries that form the staggered module. In other words,  these two primaries are coupled in the interacting theory.
Our perturbative analysis is encumbered by the fact that 
$\phi^2$ and $\partial^2 \phi^2$ have scaling dimensions that differ by an integer in the free limit.  This integer difference violates a genericity assumption of ref.\ \cite{Sen:2017gfr} where the authors study a general framework for first order perturbation theory by marginal operators in conformal field theory.   Guided by the results of the boundary bootstrap, we will see that this integer difference leads to operator mixing \cite{Berenstein:2016avf} and some novel features in the $\epsilon$ expansion.

An outline of the work is as follows.
In section \ref{sec:review}, we set up notation for boundary conformal field theory, including a review of crossing symmetry between the bulk and boundary channels. Section \ref{sec: The Free Box^2 bCFT in d=6} contains our analysis of the staggered module.  In section \ref{sec: The Interacting Box^2 bCFT in d=6-epsilon}, we analyze the interacting model (\ref{action}) in $6-\epsilon$ dimensions, taking advantage of conformal symmetry and the crossing symmetry, as well as the equation of motion, in order to reduce our reliance on Feynman diagram computations. 
We conclude in section \ref{sec:discussion} with a brief discussion of the results.

\section{Boundary Conformal Bootstrap: A Review}
\label{sec:review}

In this section, we review some basic aspects of correlators in boundary conformal field theory (bCFT) and the boundary conformal bootstrap of the bCFT data \cite{Liendo:2012hy}. We begin by introducing the notation and conventions used throughout this work. The background geometry for our theories is the Euclidean upper half-space
\begin{equation}
\mathbb{R}^d_+=\left\{x^\mu=(x_\parallel^a,x_n)\in\mathbb{R}^d: x_\parallel^a\in\mathbb{R}^{d-1},x_n\in\mathbb{R}_{\geq 0}\right\}
\end{equation}
with a boundary located at $x_n=0$. The bulk indices $\mu,\nu$ range from $1$ to $d$, while the boundary indices $a,b$ range from $1$ to $d-1$. We use the subscript $\parallel$ for quantities tangential to the boundary and the subscript $n$ for quantities normal to the boundary. For example, the Laplacian on $\mathbb{R}^d_+$ decomposes as $\Box=\Box_\parallel+\partial_n^2$.

A bCFT is a CFT defined on a space with a boundary, where a conformal boundary condition is imposed. The conformal boundary condition breaks the bulk conformal group $SO(d+1,1)$ down to the boundary conformal group $SO(d,1)$. Specifically, starting with the generators $P_\mu$, $M_{\mu\nu}$, $D$, and $K_\mu$ of the bulk conformal algebra
\begin{equation}
    \begin{split}
    & [D,P_\mu]=P_\mu,\quad [D,K_\mu]=-K_\mu,\quad [K_\mu,P_\nu]=2\delta_{\mu\nu}D-2M_{\mu\nu} \ ,\\
    & [M_{\mu\nu},P_\rho]=\delta_{\nu\rho}P_\mu-\delta_{\mu\rho}P_\nu,\quad [M_{\mu\nu},K_\rho]=\delta_{\nu\rho}K_\mu-\delta_{\mu\rho}K_\nu \ ,\\
    &[M_{\mu\nu},M_{\rho\sigma}]=\delta_{\mu\sigma}M_{\nu\rho}+\delta_{\mu\rho}M_{\sigma\nu}+\delta_{\nu\rho}M_{\mu\sigma}+\delta_{\nu\sigma}M_{\rho\mu} \ ,
    \end{split}
\end{equation}
we lose $P_n$, $M_{\mu n}$, and $K_n$. Bulk primaries are those annihilated by all $K_\mu$'s, while boundary primaries are those annihilated by all $K_a$'s. We use indices $i,j,k$ to label bulk primaries, and indices $\alpha,\beta$ to label boundary primaries. To further distinguish between bulk and boundary quantities, boundary operators and their scaling dimensions are denoted with a hat. For example, if $\phi(x)$ is a bulk operator with scaling dimension $\Delta_\phi$, we may denote the leading operator in its boundary operator expansion (BOE) as $\hat \phi(x)$, where 
\begin{equation}
    \phi(x_\parallel, x_n) \sim x_n^{\hat \Delta_\phi - \Delta_\phi} \hat \phi(x_\parallel) \ .
\end{equation}
Throughout this work, we use $\sim$ to indicate the leading-order term up to a numerical prefactor.

\subsection{Correlators in the Presence of a Boundary}

The boundary conformal symmetry $SO(d,1)$ imposes strong constraints on correlators in bCFT. Correlators of boundary operators are identical to those in a $d-1$ dimensional CFT. In other words, 
\begin{equation}
    \braket{\hat{\mathcal{O}}(x_\parallel)}=\begin{dcases}
        1 & \text{if $\hat{\mathcal{O}}$ is the boundary identity $\hat{I}$}\\
        0 & \text{otherwise}  
    \end{dcases} \ .
\end{equation}
Focusing on the scalar sector of the boundary spectrum, 
\begin{equation}
    \braket{\hat{\mathcal{O}}_\alpha(x_\parallel)\hat{\mathcal{O}}_\beta(y_\parallel)}=\frac{\delta_{\alpha\beta}}{(x_\parallel-y_\parallel)^{2\hat{\Delta}_\alpha}} \ .
\end{equation}
Correlators of bulk operators are less constrained than those in a $d$ dimensional CFT. Bulk scalar operators can get a nontrivial expectation value
\begin{equation}
    \braket{\mathcal{O}(x)}=\frac{a_\mathcal{O}}{(2x_n)^{\Delta_\mathcal{O}}} \ .
\end{equation}
One-point functions for bulk spinning operators vanish due to rotational invariance. Given two points $x$ and $y$ in the bulk, we can construct the conformally invariant cross-ratio 
\begin{equation}
    \xi=\frac{(x-y)^2}{4x_ny_n}
\end{equation}
of $\mathbb{R}_+^d$. Focusing on the scalar sector of the bulk spectrum,
\begin{equation}
    \braket{\mathcal{O}_i(x)\mathcal{O}_j(y)}=\frac{f_{ij}(\xi)}{(2x_n)^{\Delta_i}(2y_n)^{\Delta_j}}
\end{equation}
where $f_{ij}(\xi)$ is an undetermined function.

\subsection{Crossing Symmetry}

$f_{ij}(\xi)$ is further constrained by crossing symmetry. We can bring two bulk operators $\mathcal{O}_i$ and $\mathcal{O}_j$ close to each other and expand the product as an infinite sum over bulk operators
\begin{equation}
\label{eqn: OPE}
    \mathcal{O}_i(x)\mathcal{O}_j(y)=\sum_{k}\lambda_{ijk}C_{ijk}(x-y,\partial_y)\mathcal{O}_k(y) \ .
\end{equation}
This is the operator product expansion (OPE) where $\lambda_{ijk}$ is the OPE coefficient. The differential operator $C_{ijk}(x,\partial_y)$ is fixed by bulk conformal symmetry. For example, if $\mathcal{O}_i$, $\mathcal{O}_j$, and $\mathcal{O}_k$ are all scalars, then \cite{McAvity:1995zd, Meinerinotes} 
\begin{equation}
\label{Cijk}
    C_{ijk}(x,\partial_y)=x^{\Delta_k-\Delta_i-\Delta_j}\sum_{n,m=0}^\infty\frac{(-1)^m(\frac{\Delta_k+\Delta_i-\Delta_j}{2})_{m} (\frac{\Delta_k-\Delta_i+\Delta_j}{2})_{m+n}}{4^mm!n!\left(\Delta_k-\frac{d}{2}+1\right)_m(\Delta_k)_{2m+n}}(x\cdot\partial_y)^n(x^2\Box_y)^m
\end{equation}
where $(a)_l=\Gamma(a+l)/\Gamma(a)$ is the Pochhammer symbol. By evaluating
\begin{equation}
    C_{ijk}(x-y,\partial_y)\braket{\mathcal{O}_k(y)} \ ,
\end{equation}
we obtain the bulk channel decomposition of $f_{ij}(\xi)$
\begin{equation}
\label{eqn: definition of bulk channel}
    f_{ij}(\xi)=\sum_k \lambda_{ijk}a_k G_\text{bulk}(\Delta_k;\xi) \ .
\end{equation}
$G_\text{bulk}(\Delta;\xi)$ is the conformal block associated with $\mathcal{O}_k$ \cite{McAvity:1995zd}: 
\begin{equation}
\label{eqn: bulk block}
    G_\text{bulk}(\Delta;\xi)=\xi^{-\frac{1}{2}(\Delta_i+\Delta_j-\Delta)}\,
   _2F_1\left(\frac{\Delta+\Delta_i-\Delta_j }{2},\frac{\Delta-\Delta_i+\Delta_j }{2};\Delta-\frac{d}{2} +1;-\xi \right) \ .
\end{equation}

At the same time, we can bring a bulk operator $\mathcal{O}_i$ close to the boundary and expand it as an infinite sum over boundary operators:
\begin{equation}
\label{eqn: BOE}
\mathcal{O}_i(x) = \sum_{\alpha} \mu_{i\alpha} D_{i\alpha}(x_n,\partial_\parallel) \hat{\mathcal{O}}_\alpha(x_\parallel) \ .
\end{equation}
This is the BOE where $\mu_{i\alpha}$ are the BOE coefficients. The differential operator $D_{i\alpha}(x_n,\partial_\parallel)$ is fixed by boundary conformal symmetry. For example, if $\mathcal{O}_i$ is a scalar, then by boundary conformal invariance, $\hat{\mathcal{O}}_\alpha$ must be a scalar and \cite{McAvity:1995zd}
\begin{equation}
    D_{i\alpha}(x_n,\partial_\parallel)=x_n^{\hat{\Delta}_\alpha-\Delta_i}\sum_{m=0}^\infty \frac{(-1)^m}{4^mm!(\hat{\Delta}_\alpha-\frac{d-3}{2})_m}(x_n^2\Box_\parallel)^m \ .
\end{equation}
By evaluating
\begin{equation}
\label{eqn: boundary conformal block as some derivative operator on the boundary two-point correlator}
    D_{i\alpha}(x_n,\partial_{x,\parallel})D_{j\beta}(y_n,\partial_{y,\parallel})\braket{\hat{\mathcal{O}}_\alpha(x_\parallel)\hat{\mathcal{O}}_\beta(y_\parallel)} \ ,
\end{equation}
we obtain the boundary channel decomposition of $f_{ij}(\xi)$
\begin{equation}
\label{eqn: definition of boundary channel}
    f_{ij}(\xi)=\sum_\alpha\mu_{i\alpha}\mu_{j\alpha} G_\text{bry}(\hat{\Delta}_\alpha;\xi)
\ .
\end{equation}
$G_\text{bry}(\hat{\Delta}_\alpha;\xi)$ is the conformal block associated with $\hat{\mathcal{O}}_\alpha$
\cite{McAvity:1995zd, LACESnotes}:
\begin{equation}
\label{boundaryblock}
  G_\text{bry}(\hat \Delta ; \xi) =
  \left( \frac{2}{1+2 \xi}\right)^{\hat \Delta}
  {}_2F_1 \left(
 \frac{\hat \Delta}{2}, \frac{\hat \Delta+1}{2}, \hat \Delta + \frac{3-d}{2}, \frac{1}{(1+2 \xi)^2 }
  \right) \ .
\end{equation}

The equivalence of these two decompositions \eqref{eqn: definition of bulk channel} and \eqref{eqn: definition of boundary channel}  is a crossing symmetry for two-point correlators in bCFT:
\begin{equation}
\label{eqn: definition of crossing equation}
    \sum_k \lambda_{ijk}a_k G_\text{bulk}(\Delta_k;\xi)=\sum_\alpha\mu_{i\alpha}\mu_{j\alpha} G_\text{bry}(\hat{\Delta}_\alpha;\xi) \ .
\end{equation} 
A pictorial representation of the crossing symmetry is
\begin{equation}
\label{eqn: picture of crossing equation}
    \sum_k\begin{tikzpicture}[scale=0.6,baseline={([yshift=-2ex]current bounding box.center)},vertex/.style={anchor=base,
    circle,fill=black!25,minimum size=18pt,inner sep=2pt}]
    \node[circle,fill=black,inner sep=0pt,minimum size=5pt,label=above:{$\mathcal{O}_j$}] at (1.5,2) {};
    \node[circle,fill=black,inner sep=0pt,minimum size=5pt,label=above:{$\mathcal{O}_i$}] at (-1.5,2) {};
    \draw [thick] (1.5,2) -- (0,1);
    \draw [thick] (-1.5,2) -- (0,1);
    \draw [thick,blue] (0,1) -- node[right] {$\Delta_k$} (0,0);
    \draw [thick] (2,0) -- (-2,0);
    \end{tikzpicture}=\sum_\alpha\begin{tikzpicture}[scale=0.6,baseline={([yshift=-2ex]current bounding box.center)},vertex/.style={anchor=base,
    circle,fill=black!25,minimum size=18pt,inner sep=2pt}]
    \node[circle,fill=black,inner sep=0pt,minimum size=5pt,label=above:{$\mathcal{O}_j$}] at (1.5,2) {};
    \node[circle,fill=black,inner sep=0pt,minimum size=5pt,label=above:{$\mathcal{O}_i$}] at (-1.5,2) {};
    \draw [thick] (1.5,2) -- (1.5,0);
    \draw [thick] (-1.5,2) -- (-1.5,0);
    \draw [thick] (-2,0) -- (-1.5,0);
    \draw [thick] (2,0) -- (1.5,0);
    \draw [thick,red] (-1.5,0) -- node[above] {$\hat{\Delta}_\alpha$} (1.5,0);
    \end{tikzpicture} \ .
\end{equation}
Boundary conformal bootstrap attempts to solve the crossing equation \eqref{eqn: definition of crossing equation} and obtain the bCFT data $\{\Delta_i,\lambda_{ijk}a_k,\hat{\Delta}_\alpha,\mu_{i\alpha}\}$.

\section{The Free $\Box^2$ bCFT in $d=6$}
\label{sec: The Free Box^2 bCFT in d=6}

The free $\Box^2$ CFT 
\begin{equation}
\label{eqn: free action}
S_0=\frac{\kappa}{2}\int\diff^dx\,\phi\Box^2\phi
\end{equation}
is defined on $\mathbb{R}^d$.
We include the parameter 
\begin{equation}
\label{eqn: kappa factor}
\kappa \equiv \frac{1}{16 \pi^{\frac{d}{2}}} \Gamma \left( \frac{d-4}{2} \right) \xrightarrow{d \to 6} \frac{1}{16 \pi^3} \ ,
\end{equation}
so that the two-point correlator $\langle \phi(x) \phi(0) \rangle = x^{-(d-4)}$ is CFT-normalized in the absence of a boundary.
 This is a non-unitary CFT because the scaling dimension of $\phi$
\begin{equation}
\label{eqn: engineering dimension of phi}
\Delta_\phi^{(0)}=\frac{d-4}{2}
\end{equation}
violates the scalar unitarity bound $(d-2)/2$. Due to non-unitarity, there exist states that cannot be written as linear combinations of primaries and descendants in $d=6$. These states give rise to a reducible but indecomposable representation $\mathcal{H}_\text{sta}$ of the conformal algebra, which ref.\ \cite{Brust:2016gjy} referred to as an extended module. We will identify $\mathcal{H}_\text{sta}$ as a staggered module, which is mathematically described by a non-split short exact sequence.

Ref.\ \cite{Brust:2016gjy} proposed a method to construct the conformal block associated with $\mathcal{H}_\text{sta}$ by first working in $d \neq 6$ and then taking the $d \to 6$ limit. We will verify their proposal in the case of the free $\Box^2$ bCFT, using the boundary conformal bootstrap. Moreover, the identification of $\mathcal{H}_\text{sta}$ as a staggered module allows us to interpret the staggered conformal block as a generalized eigenvector of the quadratic Casimir. This interpretation provides an alternative derivation of the staggered conformal block by working directly in $d=6$.

The free $\Box^2$ bCFT is defined on $\mathbb{R}^d_+$. The action \eqref{eqn: free action} alone is not sufficient to describe a bCFT; we also need to impose a conformal boundary condition at the boundary $x_n=0$. The standard lore of conformal boundary conditions was proposed by Cardy \cite{Cardy:1984bb}: $T_{na}\rvert_{\text{bry}}=0$. Recently, ref.\ \cite{Chalabi:2022qit} introduced a different notion of conformal boundary conditions in terms of boundary primaries. It was explicitly shown that this new notion is more general than the Cardy condition in the case of the free $\Box^2$ bCFT. Furthermore, this new notion is more convenient for the boundary conformal bootstrap. We will bootstrap the $\braket{\phi\phi}_0$ correlator, where $\braket{\cdots}_0$ denotes correlators in the free theory, while $\braket{\cdots}$ is reserved for correlators in the interacting theory that will be studied in section~\ref{sec: The Interacting Box^2 bCFT in d=6-epsilon}. The result of the bootstrap agrees with the Green's function of $\Box^2$ on $\mathbb{R}^d_+$ calculated in \cite{Chalabi:2022qit}.

\subsection{The $\Box^2$ Staggered Module}
\label{subsec: Bulk Spectrum}

To prepare for the bootstrap of $\braket{\phi\phi}_0$, we study the $\phi \times \phi$ OPE, 
\begin{equation}
\phi(x)\phi(y) = \sum_{k}\lambda_k C(x-y,\partial_y)\mathcal{O}_k(y) \ .
\end{equation}
In general, any even-spin, traceless, symmetric tensor $\mathcal{O}_k$ could appear on the right-hand side. However, only $\mathcal{O}_k$ with a non-zero one-point function will contribute to $\braket{\phi\phi}_0$. By boundary conformal symmetry, $\mathcal{O}_k$ must be a scalar. Since the theory is free, $\mathcal{O}_k$ is either the identity $I$ or bilinear in $\phi$. Therefore, our task is reduced to studying scalar primaries that are bilinear in $\phi$.

The spectrum of the free $\Box^2$ CFT contains only two scalar primaries that are bilinear in $\phi$ \cite{Brust:2016gjy}:
\begin{equation}
\label{eqn: definition of d^2 phi^2}
    \phi^2,\quad \partial^2\phi^2\coloneqq\frac{1}{2}\Box\phi^2+\frac{d-6}{2}\phi\Box\phi \ .
\end{equation}
Their norms\footnote{The inner product of two states $\ket{\mathcal{O}_1}$ and $\ket{\mathcal{O}_2}$ is defined by
\begin{equation*}
    \braket{\mathcal{O}_1|\mathcal{O}_2}=\lim_{x\to 0}\braket{\mathcal{O}_1(x)^\dagger\mathcal{O}_2(x)}
\end{equation*}
where in radial quantization, the action of conjugation on a scalar primary $\mathcal{O}$ with scaling dimension $\Delta$ is
\begin{equation*}
    \mathcal{O}(x)^\dagger=x^{-2\Delta}\mathcal{O}(x_i),\quad x_i^\mu=\frac{x^\mu}{x^2} \ .
\end{equation*}
It is extended to derivatives by antilinearity
\begin{equation*}
    \left[\partial_x^n\mathcal{O}(x)\right]^\dagger=\partial_x^n[x^{-2\Delta}\mathcal{O}(x_i)] \ .
\end{equation*}
It is also extended to the normal-ordered products of $\mathcal{O}$ and their derivatives in an obvious way.} are
\begin{equation}
\label{eqn: norm of d^2 phi^2}
    \braket{\phi^2|\phi^2}=2,\quad \braket{\partial^2\phi^2|\partial^2\phi^2}=-(d-6)(d-4)^2d \ .
\end{equation}
Note the norm in this non-unitary theory is not positive definite.
In $d=6$, $\partial^2\phi^2$ becomes a descendant of $\phi^2$ and its norm vanishes.

If our CFT were unitary, primary descendants would be null, i.e.\ orthogonal to every other state in the theory. This is because the Hilbert space $\mathcal{H}$ of a unitary CFT is spanned by primaries and their descendants
\begin{equation}
\label{eqn: unitary CFT Hilbert space decomposition}
\mathcal{H}=\bigoplus_{\text{primary }\mathcal{O}}\mathcal{H}_{\mathcal{O}} \ , 
\end{equation}
where $\mathcal{H}_\mathcal{O}$ is the conformal family associated with the primary $\mathcal{O}$. 

However, the $d=6$ primary descendant $\ket{\partial^2\phi^2}$ is not null due to the breakdown of \eqref{eqn: unitary CFT Hilbert space decomposition}. To understand the connection between nullity and the decomposition of $\mathcal{H}$, consider the space of $\Delta =d-2$ scalar states that are bilinear in $\phi$. This space is two-dimensional, and $\ket{\phi\Box\phi}$ and $\ket{(\partial\phi)^2}$ constitute an orthogonal basis
\begin{equation}
    \begin{pmatrix}
        \braket{\phi\Box\phi|\phi\Box\phi}&\braket{\phi\Box\phi|(\partial\phi)^2}\\
        \braket{(\partial\phi)^2|\phi\Box\phi}&\braket{(\partial\phi)^2|(\partial\phi)^2}
    \end{pmatrix}=\begin{pmatrix}
            -4d(d-4)&0\\
            0&2d(d-4)^2
        \end{pmatrix} \ .
\end{equation}
The states $\ket{\partial^2\phi^2}$ and $\ket{\Box\phi^2}$ live in this space 
\begin{equation}
\label{eqn: change of basis}
    \begin{pmatrix}
        \ket{\partial^2\phi^2}\\
        \ket{\Box\phi^2}
    \end{pmatrix}=\begin{pmatrix}
        1&(d-4)/2\\
        2&2
    \end{pmatrix}\begin{pmatrix}
        \ket{(\partial\phi)^2}\\
        \ket{\phi\Box\phi}
    \end{pmatrix} \ .
\end{equation}
In $d\neq 6$, \eqref{eqn: change of basis} is a change of basis, so all states in this space can be written as linear combinations of the primary $\ket{\partial^2\phi^2}$ and the descendant $\ket{\Box\phi^2}$. In $d=6$, $\ket{\partial^2\phi^2}$ and $\ket{\Box\phi^2}$ do not span the space because they are not independent
\begin{equation}
    \ket{\partial^2\phi^2}=\frac{1}{2}\ket{\Box\phi^2} 
    \ . 
\end{equation}
All states in 
\begin{equation}
\label{eqn: extension operators}
    \left\{\alpha\ket{(\partial\phi)^2}+\beta\ket{\phi\Box\phi}:\alpha\neq \beta\right\}\subset\mathcal{H}
\end{equation}
cannot be written as linear combinations of primaries and descendants, so \eqref{eqn: unitary CFT Hilbert space decomposition} breaks down. The $d=6$ primary descendant $\ket{\partial^2\phi^2}$ is not null because none of the states in \eqref{eqn: extension operators} is orthogonal to $\ket{\partial^2\phi^2}$
\begin{equation}
    \alpha\neq\beta\quad\Rightarrow\quad\braket{\alpha(\partial\phi)^2+\beta\phi\Box\phi|\partial^2\phi^2}=48(\alpha-\beta)\neq 0 \ .
\end{equation}

So what is the correct description of the $d=6$ Hilbert space $\mathcal{H}$? Let's select a state in \eqref{eqn: extension operators} and denote it by $\ket{\mathcal{O}_\text{cyc}}$. Acting on $\ket{\mathcal{O}_\text{cyc}}$ with $P_\mu$'s, we can generate a family of states that cannot be written as linear combinations of primaries and descendants. Acting on $\ket{\mathcal{O}_\text{cyc}}$ with a $K_\mu$, we get a descendant of $\ket{\phi^2}$ 
\begin{equation}
    K_\mu\ket{\mathcal{O}_\text{cyc}}=2(\alpha-\beta)P_\mu\ket{\phi^2} \ .
\end{equation}

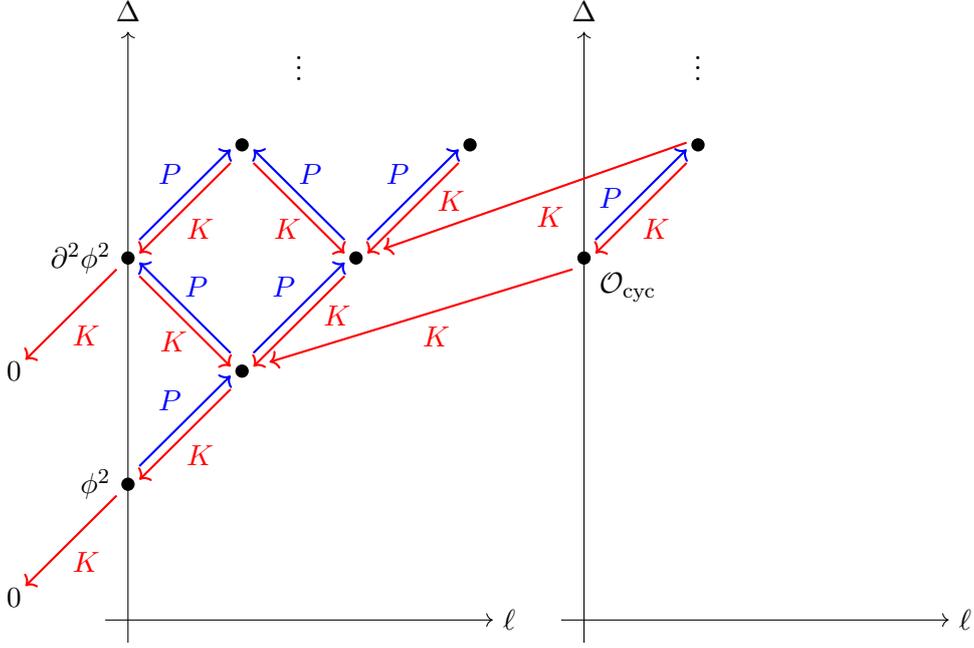
\begin{figure}[ht]
    \centering
    \begin{tikzpicture}[scale=3]
        \draw[->] (-0.1, -0.6) -- (1.6, -0.6) node[right] {$\ell$};
        \draw[->] (0, -0.7) -- (0, 2) node[above] {$\Delta$};
        \node[circle,fill=black,inner sep=0pt,minimum size=5pt,label=left:{$\phi^2$}] (phi^2) at (0,0) {};
        \node[circle,fill=black,inner sep=0pt,minimum size=5pt] (d_mu phi^2) at (0.5,0.5) {};
        \node[circle,fill=black,inner sep=0pt,minimum size=5pt] (d_mu d_nu phi^2) at (1,1) {};
        \node[circle,fill=black,inner sep=0pt,minimum size=5pt,label=left:{$\partial^2\phi^2$}] (d^2 phi^2) at (0,1) {};
        \node[circle,fill=black,inner sep=0pt,minimum size=5pt] (d_mu d_nu d_rho phi^2) at (1.5,1.5) {};
        \node[circle,fill=black,inner sep=0pt,minimum size=5pt] (d_mu d^2 phi^2) at (0.5,1.5) {};

        \node[] (0_1) at (-0.5,-0.5) {$0$};
        \node[] (0_2) at (-0.5,0.5) {$0$};

        \draw[->] (1.9, -0.6) -- (3.6, -0.6) node[right] {$\ell$};
        \draw[->] (2, -0.7) -- (2, 2) node[above] {$\Delta$};
        \node[circle,fill=black,inner sep=0pt,minimum size=5pt,label=below right:{$\mathcal{O}_\text{cyc}$}] (O_cyc) at (2,1) {};
        \node[circle,fill=black,inner sep=0pt,minimum size=5pt] (d_mu O_cyc) at (2.5,1.5) {};

        \draw [->,thick,blue] (0.05,0.08) -- node[above,xshift=-0.2cm] {$P$} (0.45,0.48);
        \draw[<-,thick,red] (0.05,0.02) -- node[below,xshift=0.2cm] {$K$} (0.45,0.42);
        \draw [->,thick,blue] (0.05,1.08) -- node[above,xshift=-0.2cm] {$P$} (0.45,1.48);
        \draw[<-,thick,red] (0.05,1.02) -- node[below,xshift=0.2cm] {$K$} (0.45,1.42);
        \draw [->,thick,blue] (0.55,0.58) -- node[above,xshift=-0.2cm] {$P$} (0.95,0.98);
        \draw[<-,thick,red] (0.55,0.52) -- node[below,xshift=0.5cm,yshift=0.35cm] {$K$} (0.95,0.92);
        \draw [->,thick,blue] (1.05,1.08) -- node[above,xshift=-0.2cm] {$P$} (1.45,1.48);
        \draw[<-,thick,red] (1.05,1.02) -- node[below,xshift=0.5cm,yshift=0.37cm] {$K$} (1.45,1.42);

        \draw [->,thick,blue] (0.45,0.58) -- node[above,xshift=0.15cm] {$P$} (0.05,0.98);
        \draw[->,thick,red] (0.05,0.92) -- node[below,xshift=-0.15cm] {$K$} (0.45,0.52);
        \draw [->,thick,blue] (0.95,1.08) -- node[above,xshift=0.15cm] {$P$} (0.55,1.48);
        \draw[->,thick,red] (0.55,1.42) -- node[below,xshift=-0.15cm] {$K$} (0.95,1.02);
        
        \draw[<-,thick,red] (-0.45,-0.45) -- node[below,xshift=0.2cm] {$K$} (-0.05,-0.05);
        \draw[<-,thick,red] (-0.45,0.55) -- node[below,xshift=0.2cm] {$K$} (-0.05,0.95);

        \draw [->,thick,blue] (2.05,1.08) -- node[above,xshift=-0.4cm,yshift=-0.32cm] {$P$} (2.45,1.48);
        \draw[<-,thick,red] (2.05,1.02) -- node[below,xshift=0.2cm] {$K$} (2.45,1.42);
        \draw[->,thick,red] (1.95,0.95) -- node[below,xshift=0.2cm] {$K$} (0.62,0.54);
        \draw[->,thick,red] (2.45,1.51) -- node[below,xshift=0.2cm] {$K$} (1.12,1.04);

        \node[] (dots_1) at (0.75,2.0) {};
        \node[] (dots_2) at (0.75,1.75) {};
        \path (dots_1) -- node[auto=false]{\vdots} (dots_2);

        \node[] (dots_3) at (2.5,2.0) {};
        \node[] (dots_4) at (2.5,1.75) {};
        \path (dots_3) -- node[auto=false]{\vdots} (dots_4);
    \end{tikzpicture}
    \caption{The $\Box^2$ staggered module $\mathcal{H}_\text{sta}$ in $d=6$.}
    \label{fig: d=6 staggered module}
\end{figure}

From figure~\ref{fig: d=6 staggered module}, we see that the $d=6$ Hilbert space $\mathcal{H}$ can be described by replacing $\mathcal{H}_{\phi^2}\oplus\mathcal{H}_{\partial^2\phi^2}$ in \eqref{eqn: unitary CFT Hilbert space decomposition} with 
\begin{equation}
\label{eqn: staggered module}
    \mathcal{H}_\text{sta}= \Span\left\{\ket{\phi^2},P_\mu \ket{\phi^2},\cdots,\ket{\mathcal{O}_\text{cyc}},P_\mu\ket{\mathcal{O}_\text{cyc}},\cdots\right\} \ .
\end{equation}
Note that $K^2\ket{\mathcal{O}_\text{cyc}}=48(\alpha-\beta)\ket{\phi^2}$. Therefore, $\ket{\mathcal{O}_\text{cyc}}$ is a cyclic vector for $\mathcal{H}_\text{sta}$, meaning that the action of the conformal algebra on it generates $\mathcal{H}_\text{sta}$.

As a representation of the conformal algebra, $\mathcal{H}_\text{sta}$ is reducible: $\mathcal{H}_{\phi^2}$ and $\mathcal{H}_{\partial^2\phi^2}$ are its proper subrepresentations. However, it is not decomposable into irreducible representations. To better characterize $\mathcal{H}_\text{sta}$, we study the action of the quadratic Casimir
\begin{equation}
    \mathcal{C}=D^2-\frac{P^\mu K_\mu+K^\mu P_\mu}{2}-\frac{1}{2}M^{\mu\nu}M_{\mu\nu}
\end{equation}
on $\mathcal{H}_\text{sta}$. Recall that $\mathcal{C}$ acts with the same eigenvalue $\Delta(\Delta-d)$ on every state in the conformal family $\mathcal{H}_\mathcal{O}$ of a scalar primary $\mathcal{O}$ with scaling dimension $\Delta$.

We arrange the basis vectors for $\mathcal{H}_\text{ext}$ in the following order
\begin{equation}
\ket{\phi^2},\, P_\mu\ket{\phi^2},\, P_\mu P_\nu\ket{\phi^2},\, P_\mu P_\nu P_\rho\ket{\phi^2},\,\cdots,\, (\ket{\partial^2\phi^2},\ket{\mathcal{O}_\text{cyc}}),\, (P_\mu\ket{\partial^2\phi^2}, P_\mu\ket{\mathcal{O}_\text{cyc}}),\,\cdots
\end{equation}
where $\ket{\partial^2\phi^2}=\frac{1}{2}P^2\ket{\phi^2}$. Denote a string of $m$ $P_\mu$'s with no contractions schematically as $P_{\mu_1}\cdots P_{\mu_m}$. The states $P_{\mu_1}\cdots P_{\mu_m}\ket{\phi^2}$ are eigenvectors of $\mathcal{C}$
\begin{equation}
\label{eqn: action of Casimir 1}
\mathcal{C}P_{\mu_1}\cdots P_{\mu_m}\ket{\phi^2}=-8P_{\mu_1}\cdots P_{\mu_m}\ket{\phi^2} \ .
\end{equation}
Denote a generic string of $m$ $P_\mu$'s schematically as $(P)^m$. Each pair $$\left((P)^m\ket{\partial^2\phi^2},(P)^m\ket{\mathcal{O}_\text{cyc}}\right)$$ generates a Jordan block of rank 2
\begin{align}
\begin{split}
\label{eqn: action of Casimir 2}
\mathcal{C}(P)^m\ket{\partial^2\phi^2}&=-8(P)^m\ket{\partial^2\phi^2}, \\
\mathcal{C}(P)^m\ket{\mathcal{O}_\text{cyc}}&=-4(\alpha-\beta) (P)^m\ket{\partial^2\phi^2}-8(P)^m\ket{\mathcal{O}_\text{cyc}} \ .
\end{split}
\end{align}
Hence, $\mathcal{C}$ has a Jordan normal form on $\mathcal{H}_\text{sta}$
\begin{equation}
\mathcal{C}=\begin{pNiceMatrix}
-8 & 0 & 0 & 0 & 0 & 0 & 0 & 0\\
0 & \ddots & 0 & 0 & 0 & 0 & 0 & 0\\
0 & 0 & -8 & 0 & 0 & 0 & 0 & 0\\
0 & 0 & 0 & \Block[borders={bottom,right,top,left}]{2-2}{} -8 & -4(\alpha-\beta) & 0 & 0 & 0\\
0 & 0 & 0 & 0 & -8 & 0 & 0 & 0\\
0 & 0 & 0 & 0 & 0 & \ddots & 0 & 0\\
0 & 0 & 0 & 0 & 0 & 0 & \Block[borders={top,left}]{2-2}{} -8 & -4(\alpha-\beta) \\
0 & 0 & 0 & 0 & 0 & 0 & 0 & -8\\
\end{pNiceMatrix}
\end{equation}
Note that the restriction of $\mathcal{C}+8$ on $\mathcal{H}_\text{sta}$ is nilpotent
\begin{equation}
\label{eqn: (C+8)^2 is nilpotent}
    \ker(\mathcal{C}+8)^2=\mathcal{H}_\text{sta} \ .
\end{equation}
This observation allows us to interpret the conformal block associated with $\mathcal{H}_\text{sta}$ as a rank-2 generalized eigenvector of $\mathcal{C}$. In section~\ref{subsec: The Staggered Conformal Block}, we will solve the generalized Casimir equation and recover the staggered conformal block proposed by \cite{Brust:2016gjy}.

Let $\iota\colon \mathcal{H}_{\phi^2}\to \mathcal{H}_\text{sta}$ be the inclusion. Eqs.~\eqref{eqn: action of Casimir 1} and \eqref{eqn: action of Casimir 2} imply that $\pi=\mathcal{C}+8$ is a surjection from $\mathcal{H}_\text{sta}$ onto $\mathcal{H}_{\partial^2\phi^2}$ and
\begin{equation}
    \ker\pi=\mathcal{H}_{\phi^2}=\Im\iota \ .
\end{equation}
Thus, $\mathcal{H}_\text{sta}$ can be described by a short exact sequence
\begin{equation}
    0\longrightarrow \mathcal{H}_{\phi^2}\overset{\iota}{\longrightarrow}\mathcal{H}_\text{sta}\overset{\pi}{\longrightarrow}\mathcal{H}_{\partial^2\phi^2}\longrightarrow 0 \ .
\end{equation}
This short exact sequence is non-split $\mathcal{H}_\text{sta}\ncong\mathcal{H}_{\phi^2}\oplus\mathcal{H}_{\partial^2\phi^2}$. Mathematically, $\mathcal{H}_\text{sta}$ is known as a staggered module \cite{Rohsiepe:1996qj}.

\subsection{Conformal Boundary Conditions}
\label{subsec: Boundary Spectrum}

To render the free $\Box^2$ bQFT a bCFT, we will study the possible conformal boundary conditions. Additionally, for a given conformal boundary condition, we want to determine which boundary primaries $\hat{\mathcal{O}}_\alpha$ will appear in the the BOE of $\phi$
\begin{equation}
    \phi(x)=\sum_{a}\mu_{a}D(x_n,\partial_{\parallel})\hat{\mathcal{O}}_a(x_\parallel) \ .
\end{equation}
By boundary conformal invariance, $\hat{\mathcal{O}}_\alpha$ must be a scalar. Since the theory is free and possesses $\mathbb{Z}_2$ symmetry, $\hat{\mathcal{O}}_\alpha$ must be linear in $\hat{\phi}(x_\parallel)$.

From the bulk primary $\ket{\phi}$, we can construct boundary primaries of the schematic form $((P_n)^m + \cdots)\ket{\phi}$, each of which is annihilated by all $K_a$'s but not necessarily by $K_n$. It was shown in ref.\ \cite{Chalabi:2022qit} that the boundary spectrum of the free $\Box^2$ bCFT contains only four scalar primaries that are linear in $\hat{\phi}$
\begin{equation}
\label{eqn: boundary primaries}
    \Phi^{(0)}\coloneqq\phi,\quad \Phi^{(1)}\coloneqq \partial_n\phi,\quad \Phi^{(2)}\coloneqq (\partial_n^2-\Box_\parallel)\phi,\quad \Phi^{(3)}\coloneqq(\partial_n^3+3\partial_n\Box_\parallel)\phi \ .
\end{equation}
In terms of these boundary primaries, the variation of the action \eqref{eqn: free action} can be written as
\begin{equation}
\begin{split}
    \delta S_0&=\int_{\mathbb{R}_+^d}\diff^dx\,\delta\phi\, \Box^2\phi\\
    +&\frac{1}{2}\int_{x_n=0}\diff^{d-1}x_\parallel\,\left(\Phi^{(0)}\delta\Phi^{(3)}-\Phi^{(1)}\delta\Phi^{(2)}+\Phi^{(2)}\delta\Phi^{(1)}-\Phi^{(3)}\delta\Phi^{(0)}\right) \ .
\end{split}
\end{equation}
To eliminate the boundary variation, there are four possibilities
\begin{align}
\label{eqn: conformal boundary conditions}
    \begin{split}
        &DD:\quad\Phi^{(0)}=\Phi^{(1)}=0 \ , \\
        &DN:\quad\Phi^{(0)}=\Phi^{(2)}=0 \ , \\
        &ND:\quad\Phi^{(1)}=\Phi^{(3)}=0 \ , \\
        &NN:\quad\Phi^{(2)}=\Phi^{(3)}=0 \ .
    \end{split}
\end{align} 
It was checked in \cite{Chalabi:2022qit} that the Cardy conformal boundary condition $T_{na}\rvert_{x_n=0}=0$ is satisfied if one of the $DD$, $DN$, $ND$ boundary conditions is imposed. If the $NN$ boundary condition is imposed, then
\begin{equation}
    T_{na}\rvert_{x_n=0}=-\partial^b\tau_{ba}
\end{equation}
where $\tau_{ab}$ is some purely boundary contribution to the stress tensor that is traceless and symmetric although not conserved. In this case, the generators of the boundary conformal algebra were still checked to be conserved
\begin{equation}
    \dv{P^a}{t}=\dv{M^{ab}}{t}=\dv{D}{t}=\dv{K^a}{t}=0 \ .
\end{equation}
Therefore, \eqref{eqn: conformal boundary conditions} provides four possible conformal boundary conditions, and the $NN$ boundary condition is more general than the Cardy condition.

To conclude, for a boundary primary $\hat{\mathcal{O}}_\alpha$ to appear in the BOE of $\phi$, it must be one of the $\Phi^{(i)}$'s in \eqref{eqn: boundary primaries}. Each of the conformal boundary conditions \eqref{eqn: conformal boundary conditions} removes two $\Phi^{(i)}$'s from the boundary spectrum. For example, if the $NN$ boundary condition is imposed, then $\Phi^{(2)}$ and $\Phi^{(3)}$ are removed. Consequently, only the boundary primaries $\Phi^{(0)}$ and $\Phi^{(1)}$ will appear in the BOE of $\phi$.

\subsection{The Staggered Conformal Block}
\label{subsec: The Staggered Conformal Block}

In this section, we study the conformal block associated with $\mathcal{H}_\text{sta}$ in the $d=6$ free $\Box^2$ bCFT. Ref.\ \cite{Brust:2016gjy} proposed that the staggered conformal block can be obtained by first summing the $\phi^2$ and $\partial^2\phi^2$ conformal blocks in $d \neq 6$, weighted by their OPE coefficients, and then taking the $d \to 6$ limit. We will verify their proposal by bootstrapping
\begin{equation}
\braket{\phi(x)\phi(y)}_0=\frac{f_0(\xi)}{(4x_ny_n)^{\Delta_{\phi}^{(0)}}}
\end{equation}
in the $d \neq 6$ free $\Box^2$ bCFT. We will also provide an alternative derivation of the staggered conformal block by working directly in $d=6$ and solving the generalized Casimir equation.

\subsubsection*{From the Boundary Conformal Bootstrap}

In this section, we bootstrap the free bCFT data $\{\Delta^{(0)}, \lambda a^{(0)}, \hat{\Delta}^{(0)}, \mu^{(0)}\}$. These quantities will acquire anomalous corrections in the interacting theory:
\begin{equation}
\label{eqn: general ansatz for bCFT data}
    \begin{aligned}
        \Delta&=\Delta^{(0)}+\delta\Delta^{(1)}\epsilon+O(\epsilon^2),   &\quad   \lambda a&=\lambda a^{(0)}+\delta \lambda a^{(1)}\epsilon+O(\epsilon^2),\\
        \hat{\Delta}&=\hat{\Delta}^{(0)}+\delta\hat{\Delta}^{(1)}\epsilon+O(\epsilon^2),   &\quad   \mu&=\mu^{(0)}+\delta\mu^{(1)}\epsilon+O(\epsilon^2)
\end{aligned}
\end{equation}
which will be explored in section~\ref{sec: The Interacting Box^2 bCFT in d=6-epsilon}. To simplify notation, we will omit all $(0)$ superscripts in this section, but the results here will be referenced with a $(0)$ superscript in section~\ref{sec: The Interacting Box^2 bCFT in d=6-epsilon}.

From the discussion in section~\ref{subsec: Bulk Spectrum}, we know that in $d\neq 6$, only bulk primaries $I$, $\phi^2$, and $\partial^2\phi^2$ will contribute to $\braket{\phi\phi}_0$ through the bulk channel
\begin{equation}
\label{eqn: tree-level bulk channel}
    f_0(\xi)=G_\text{bulk}(0;\xi)+\lambda a_{\phi^2}G_\text{bulk}(\Delta_{\phi^2};\xi)+\lambda a_{\partial^2\phi^2}G_\text{bulk}(\Delta_{\partial^2\phi^2};\xi)
\end{equation}
where
\begin{equation}
    G_\text{bulk}(\Delta;\xi)=\xi^{-\Delta_\phi+\frac{\Delta}{2}}\,
   _2F_1\left(\frac{\Delta }{2},\frac{\Delta }{2};\Delta-\frac{d}{2} +1;-\xi \right)
\end{equation}
and
\begin{equation}
    \Delta_{\phi^2}=d-4,\quad \Delta_{\partial^2\phi^2}=d-2 \ .
\end{equation}
From the discussion in section~\ref{subsec: Boundary Spectrum}, we know that depending on the conformal boundary condition \eqref{eqn: conformal boundary conditions} we impose, only two out of the four boundary primaries \eqref{eqn: boundary primaries} will contribute to $\braket{\phi\phi}_0$ through the boundary channel. Before specializing to a particular conformal boundary condition, we start with the most general boundary channel
\begin{equation}
    \label{eqn: tree-level boundary channel}
    f_0(\xi)= \sum_{i=0}^3 \mu_i^2 G_\text{bry}(\hat{\Delta}_i;\xi)
\end{equation}
where
\begin{equation}
    \hat{\Delta}_i =\frac{d-4}{2} + i \ .
\end{equation}
Equating the bulk channel \eqref{eqn: tree-level bulk channel} and the boundary channel \eqref{eqn: tree-level boundary channel}, we get the general crossing equation
\begin{equation}
\label{eqn: crossing equation}
    G_\text{bulk}(0;\xi)+\lambda a_{\phi^2} G_\text{bulk}(\Delta_{\phi^2};\xi)+\lambda a_{\partial^2\phi^2}G_\text{bulk}(\Delta_{\partial^2\phi^2};\xi)
= \sum_{i=0}^3 \mu_i^2 G_\text{bry}(\hat{\Delta}_i;\xi) \ .
\end{equation}
This general crossing equation \eqref{eqn: crossing equation} is satisfied provided
\begin{align}
    \begin{split}
         d(d-2)(d-4)\mu_0^2-192\mu_3^2&=2d(d-2)(d-4) \ , \\
         2(d-2)\mu_1^2+8\mu_2^2&=(d-2)(d-4) \ , \\
         \lambda a_{\phi^2}&=\frac{d}{4}\mu_0^2+\mu_1^2-\frac{d-2}{2} \ , \\
         \lambda a_{\partial^2\phi^2}&=\frac{3(d-4)}{2(d-6)}\mu_0^2+\frac{2}{d-6}\mu_1^2-\frac{2(d-4)}{d-6} \ ,
    \end{split}
\end{align}
leaving a two parameter family of solutions.  Imposing the 
$DD$, $DN$, $ND$, or $NN$ conformal boundary conditions selects out four special solutions
listed in table~\ref{tab: The free Box^2 bCFT data for the four conformal boundary conditions}.

\begin{table}[h]
\begin{center}
{\renewcommand{\arraystretch}{1.5} 
\begin{tabular}{ |c | c | c | c | c | }
\hline 
 & $DD$ & $DN$ & $ND$ & $NN$ \\ \hline                        
$\lambda a_{\phi^2}$ & $-\frac{d-2}{2}$ & $-1$ & $1$ & $\frac{d-2}{2}$ \\
$\lambda a_{\partial^2\phi^2}$ & $-\frac{2(d-4)}{d-6}$ & $-\frac{d-4}{d-6}$ & $\frac{d-4}{d-6}$ & $\frac{2(d-4)}{d-6}$ \\
$\mu_0^2$ & $0$ & $0$ & $2$ & $2$ \\
$\mu_1^2$ & $0$ & $\frac{d-4}{2}$ & $0$ & $\frac{d-4}{2}$ \\
$\mu_2^2$ & $\frac{(d-2)(d-4)}{8}$ & $0$ & $\frac{(d-2)(d-4)}{8}$ & $0$ \\
$\mu_3^2$ & $-\frac{d(d-2)(d-4)}{96}$ & $-\frac{d(d-2)(d-4)}{96}$ & $0$ & $0$ \\
\hline  
\end{tabular}
}
\end{center}
\caption{The free $\Box^2$ bCFT data for the four conformal boundary conditions.}
\label{tab: The free Box^2 bCFT data for the four conformal boundary conditions}
\end{table}

Using these bCFT data, we can construct $\braket{\phi\phi}_0$ either through the bulk channel or the boundary channel
\begin{equation}
\label{eqn: correlator solution to the crossing equation}
    f_0(\xi)
    =\begin{dcases}
    \xi^{-\frac{d-4}{2}}-\frac{d-4}{2}(1+\xi)^{-\frac{d-2}{2}}-(1+\xi)^{-\frac{d-4}{2}} & DD\\
    \xi^{-\frac{d-4}{2}}-(1+\xi)^{-\frac{d-4}{2}} & DN\\
    \xi^{-\frac{d-4}{2}}+(1+\xi)^{-\frac{d-4}{2}} & ND\\
       \xi^{-\frac{d-4}{2}}+\frac{d-4}{2}(1+\xi)^{-\frac{d-2}{2}}+(1+\xi)^{-\frac{d-4}{2}} & NN 
    \end{dcases}
\end{equation}
Eq.~\eqref{eqn: correlator solution to the crossing equation} agrees with the Green's function of $\Box^2$ on $\mathbb{R}_+^d$ calculated in \cite{Chalabi:2022qit}.

The OPE coefficient\footnote{For the remainder of this work, we will refer to $\lambda a$ as the OPE coefficient. While a slight abuse of notation --- since $\lambda a$ is actually the product of the OPE coefficient $\lambda$ and the one-point coefficient $a$ --- it is the natural bCFT data to study in the context of boundary conformal bootstrap. For the same reason, we will sometimes refer to $\mu^2$ as the BOE coefficient.} $\lambda a_{\partial^2\phi^2}$ for all conformal boundary conditions exhibits a pole at $d=6$, signaling the breakdown of the $\phi\times\phi$ OPE in $d=6$. This breakdown is attributed to the $d=6$ staggered module $\mathcal{H}_\text{sta}$, within which there exist states like $\ket{\mathcal{O}_\text{cyc}}$ that cannot be expressed as linear combinations of primaries and descendants. More explicitly, $\phi^2$ contributes to the $\phi\times\phi$ OPE as follows
\begin{equation}
\label{eqn: phi^2 contribution to the phi phi OPE}
\phi(x)\phi(0)\supset \left(1+\frac{1}{2}x\cdot\partial+\frac{d-2}{8(d-3)}(x\cdot\partial)^2 -\frac{d-4}{8(d-3)(d-6)}x^2\Box+\cdots\right)\phi^2(0) \ ,
\end{equation}
while $\partial^2\phi^2$ contributes as follows
\begin{equation}
\label{eqn: d^2phi^2 contribution to the phi phi OPE}
\phi(x)\phi(0)\supset\frac{x^2}{d(d-6)}(1+\cdots)\partial^2\phi^2(0)=\frac{x^2}{2d(d-6)}(1+\cdots)\Box\phi^2(0)+\frac{x^2}{2d}(1+\cdots)\phi\Box\phi(0) \ .
\end{equation}
(These expressions may be checked by contracting both sides with the operators $\phi^2$ and $\partial^2 \phi^2$ respectively
or derived by using (\ref{Cijk}).)
Away from $d=6$, these two contributions are well-defined independently. In $d=6$, $\partial^2\phi^2$ becomes a descendant of $\phi^2$, so we expect the contribution of $\partial^2\phi^2$ to be absorbed into the contribution of $\phi^2$. This leads us to consider the sum of the two contributions as a whole
\begin{equation}
\label{eqn: extended module contribution to the phi phi OPE}
    \phi(x)\phi(0)\supset\left(1+\frac{1}{2}x\cdot \partial+\frac{1}{6}(x\cdot\partial)^2-\frac{1}{36}x^2\Box+\cdots\right)\phi^2(0)+\frac{x^2}{12}(1+\cdots)\phi\Box\phi(0) \ .
\end{equation}
Firstly, note that the $d=6$ poles in \eqref{eqn: phi^2 contribution to the phi phi OPE} and \eqref{eqn: d^2phi^2 contribution to the phi phi OPE} precisely cancel each other out. This cancellation is expected because the free $\Box^2$ CFT is well-defined in $d=6$. Secondly, observe that the absorption is not complete, leaving behind the $\phi\Box\phi$ term in \eqref{eqn: extended module contribution to the phi phi OPE}. $\ket{\phi\Box\phi}$ lives in the space \eqref{eqn: extension operators}, so it can be taken to be $\ket{\mathcal{O}_\text{cyc}}$. Therefore, \eqref{eqn: extended module contribution to the phi phi OPE} aligns with figure~\ref{fig: d=6 staggered module}: the $\partial^2\phi^2$ conformal family is absorbed into the $\phi^2$ conformal family, while $\mathcal{O}_\text{cyc}$ and its descendants remain.

For the remainder of section~\ref{sec: The Free Box^2 bCFT in d=6}, we will focus on the $NN$ boundary condition. (A parallel analysis can be done for the $DD$, $DN$, and $ND$ boundary conditions without difficulty.) The operator $\phi^2$ contributes to the bulk channel as follows
\begin{equation}
    f_0(\xi)\supset\lambda a_{\phi^2}G_\text{bulk}(\Delta_{\phi^2};\xi)=\frac{d-2}{2}\frac{1}{(1+\xi)^{\frac{d-2}{2}}}\left(1-\frac{2\xi}{d-6}\right) \ , 
\end{equation}
while $\partial^2\phi^2$ contributes as 
\begin{equation}
    f_0(\xi)\supset\lambda a_{\partial^2\phi^2}G_\text{bulk}(\Delta_{\partial^2\phi^2};\xi)=\frac{2(d-4)}{d-6}\frac{\xi}{(1+\xi)^{\frac{d-2}{2}}} \ .
\end{equation}
Note that the $\phi^2$ conformal block has a pole at $d=6$, and its residue is proportional to the $\partial^2\phi^2$ conformal block
\begin{equation}
\label{eqn: residue of the phi^2 block in the free theory}
    \lim_{d\to 6}(d-6) G_\text{bulk}(\Delta_{\phi^2};\xi)=-\frac{2 \xi}{(1+\xi)^2}=-2\left.G_\text{bulk}(\Delta_{\partial^2\phi^2};\xi)\right\rvert_{d=6} \ ,
\end{equation}
as expected from the fact that $\partial^2\phi^2$ becomes a primary descendant of $\phi^2$ in $d=6$ \cite{Penedones:2015aga}. In $d=6$, it only makes sense to talk about these two contributions as a whole, which can be thought of as the contribution from the staggered module $\mathcal{H}_\text{sta}$
\begin{equation}
\label{eqn: extended block}
    f_0(\xi)\supset G_\text{sta}(\xi)\coloneqq\lambda a_{\phi^2}G_\text{bulk}(\Delta_{\phi^2};\xi)+\lambda a_{\partial^2\phi^2}G_\text{bulk}(\Delta_{\partial^2\phi^2};\xi)=\frac{1}{(1+\xi)^2}+\frac{1}{1+\xi} \ .
\end{equation}
This method of constructing the staggered conformal block was proposed by \cite{Brust:2016gjy}.

\subsubsection*{From the Generalized Casimir Equation}

Alternatively, we can derive the staggered conformal block $G_\text{sta}(\xi)$ by working directly in $d=6$ and solving a generalized Casimir equation. We first review the Casimir equation, deferring its generalization until slightly later.

Consider the bulk channel contribution to $\braket{\phi\phi}$ from the conformal family $\mathcal{H}_\mathcal{O}$ of a scalar primary $\mathcal{O}\ni\phi\times\phi$ with dimension $\Delta$:
\begin{equation}
    \braket{\phi(x)\phi(y)}\supset 
    \frac{\lambda a_\mathcal{O}G_\text{bulk}(\Delta;\xi)}{(4x_ny_n)^{\Delta_\phi}} \ .
\end{equation}
All the members of this family are eigenfunctions of the Casimir operator ${\mathcal C}$ with the same eigenvalue $\Delta(\Delta-d)$.
Using embedding space or other methods, it is an exercise to show that the action of the Casimir can be represented as a second order differential operator
\begin{equation}
\label{eqn: Casimir equation}
0 = \left({\mathfrak C} - \Delta(\Delta-d)\right) \left(\xi^{\Delta_{\phi}} G_{\rm bulk} (\Delta ; \xi)\right) \ ,
\end{equation}
where \cite{Liendo:2012hy}
\begin{equation}
    \mathfrak{C}=4\xi^2(1+\xi)\pdv[2]{}{\xi}-2\xi(d-2-2\xi)\pdv{}{\xi} \ .
\end{equation}
The differential equation
$\left({\mathfrak C} - \Delta (\Delta-d)\right) F(\Delta; \xi) = 0$ has a general solution
\begin{equation}
\label{eqn: general solution to the Casimir equation}
    \begin{split}
        F(\Delta;\xi)&=a_1\xi^{\frac{\Delta}{2}}\,
   _2F_1\left(\frac{\Delta}{2},\frac{\Delta}{2};\Delta-\frac{d}{2}+1;-\xi \right)\\
   &+a_2\xi^{\frac{d-\Delta}{2}}\, _2F_1\left(\frac{d-\Delta}{2},\frac{d-\Delta}{2};-\Delta+\frac{d}{2}+1;-\xi \right)\ .
    \end{split}
\end{equation}
We identify the $a_1$ solution as the conformal block associated with $\mathcal{O}$
\begin{equation}
    G_\text{bulk}(\Delta;\xi)=\xi^{-\Delta_\phi+\frac{\Delta}{2}}\,
   _2F_1\left(\frac{\Delta}{2},\frac{\Delta}{2};\Delta-\frac{d}{2}+1;-\xi \right)
\end{equation}
because it has the desired $\xi\to 0$ behavior 
\begin{equation}
\label{eqn: xi to 0 behavior of the O conformal block}
    G_\text{bulk}(\Delta;\xi)\sim \xi^{-\Delta_\phi+\frac{\Delta}{2}}
\end{equation}
required by the OPE \cite{McAvity:1995zd}. The $a_2$ solution is interpreted as the conformal block associated with the shadow operator 
\begin{equation}
\label{eqn: shadow transform}
    \Tilde{\mathcal{O}}(x)=\frac{\Gamma(\Delta)}{\Gamma(\Delta-\frac{d}{2})}\int\diff^dy\,\frac{\mathcal{O}(y)}{(x-y)^{2(d-\Delta)}}
\end{equation}
of $\mathcal{O}$ \cite{Ferrara:1972uq, Simmons-Duffin:2012juh}. $\Tilde{\mathcal{O}}$ transforms as a scalar primary with scaling dimension $\Tilde{\Delta}=d-\Delta$.

In the special case $\Tilde{\Delta}=\Delta+2n$, where $n=1,2,\cdots$, 
\begin{equation}
\label{eqn: Delta for type 3 primary descendant}
    \Delta=\frac{d}{2}-n \ , 
\end{equation}
the integral kernel of the shadow transform \eqref{eqn: shadow transform} becomes ultra-local (see p 74 of \cite{gelfand1964generalized}):
\begin{equation}
    \left.\frac{1}{\Gamma(\Delta-\frac{d}{2})}\frac{1}{x^{2(d-\Delta)}}\right\rvert_{\Delta=\frac{d}{2}-n}=\frac{(-1)^n\pi^\frac{d}{2}}{2^{2n}\Gamma(\frac{d}{2}+n)}\Box^n\delta^{(d)}(x) \ .
\end{equation}
The shadow $\Tilde{\mathcal{O}}\propto \Box^n\mathcal{O}$ is a primary descendant of $\mathcal{O}$. Indeed, following the notations of \cite{Penedones:2015aga}, \eqref{eqn: Delta for type 3 primary descendant} is the condition for $\mathcal{O}$ to have a type III primary descendant. The $\mathcal{O}$ conformal block $G_\text{bulk}(\Delta;\xi)$ then has simple poles at \eqref{eqn: Delta for type 3 primary descendant}. After regularizing it by changing the normalization  
\begin{equation}
\label{eqn: solution 1 to the Casimir equation}
    \left.\frac{1}{\Gamma(1-n)}G_\text{bulk}(\Delta;\xi)\right\rvert_{\Delta=\frac{d}{2}-n}=\frac{(-1)^n}{\Gamma(1+n)}\left.\left(\frac{\Gamma(\Tilde{\Delta}/2)}{\Gamma(\Delta/2)}\right)^2G_\text{bulk}(\Tilde{\Delta};\xi)\right\rvert_{\Delta=\frac{d}{2}-n} \ ,
\end{equation}
it becomes degenerate with the shadow conformal block $G_\text{bulk}(\Tilde{\Delta};\xi)$. The second independent solution to \eqref{eqn: Casimir equation} should have the same $\xi\to 0$ behavior \eqref{eqn: xi to 0 behavior of the O conformal block} as the $\mathcal{O}$ conformal block, with a subleading logarithmic contribution \cite{Abramowitz}.

In the $d=6$ free $\Box^2$ bCFT, the bulk primaries $\phi^2$ and $\partial^2\phi^2$ form such a shadow pair $\Tilde{\phi^2}\propto\partial^2\phi^2$. To capture the staggered module $\mathcal{H}_\text{sta}\supset\mathcal{H}_{\phi^2}\supset\mathcal{H}_{\partial^2\phi^2}$ contribution to $\braket{\phi\phi}_0$
\begin{equation}
    \braket{\phi(x)\phi(y)}_0\supset 
    \frac{G_\text{sta}(\xi)}{4x_ny_n} \ , 
\end{equation}
we have to modify the analysis above. The observation \eqref{eqn: (C+8)^2 is nilpotent} that $(\mathcal{C}+8)^2$ is nilpotent on $\mathcal{H}_\text{sta}$ implies
\begin{equation}
\label{eqn: generalized Casimir equation}
    \begin{split}
        0
        =(\mathfrak{C}+8)^2(\xi G_\text{sta}(\xi)) \ .
    \end{split}
\end{equation}
We call \eqref{eqn: generalized Casimir equation} a generalized Casimir equation because it allows us to interpret the staggered conformal block $G_\text{sta}(\xi)$ as a rank-2 generalized eigenvector of the quadratic Casimir differential operator $\mathfrak{C}$. The general solution to \eqref{eqn: generalized Casimir equation} is
\begin{equation}
    G_\text{sta}(\xi)=a_1\frac{1}{1+\xi}+a_2\frac{1}{(1+\xi)^2}+a_3 \frac{\log (\xi )}{(1+\xi)}+a_4\frac{\log (\xi )}{(1+\xi)^2} \ .
\end{equation}
The integration constants $a_i$'s can be fixed by the staggered module contribution \eqref{eqn: extended module contribution to the phi phi OPE} to the $\phi\times\phi$ OPE. Without loss of generality, we specialize to the configuration $x=(0,x_n)$ and $y=(0,y_n)$. Define $
\delta_n=x_n-y_n$, so that $\xi=\delta_n^2/(4(y_n+\delta_n)y_n)$.
Note that in $d=6$,
\begin{equation}
    \braket{\phi^2(y)}_0=\frac{1}{2y_n^2},\quad \braket{\phi\Box\phi(y)}_0=-\frac{5}{4y_n^4}
\end{equation}
when the $NN$ boundary condition is imposed. Eq.~\eqref{eqn: extended module contribution to the phi phi OPE} becomes
\begin{equation}
\label{eqn: determination of the staggered block integration constants 1}
    \braket{\phi(x)\phi(y)}_0\supset \frac{1}{2y_n^2}-\frac{\delta_n}{2y_n^3}+\frac{5\delta_n^2}{16y_n^4}+O(\delta_n^3) \ .
\end{equation}
By matching \eqref{eqn: determination of the staggered block integration constants 1} with the expansion of $G_\text{sta}(\xi)/(4 (y_n+\delta_n) y_n)$ around $\delta_n=0$, we get
\begin{equation}
    a_1=1,\quad a_2=1,\quad a_3=a_4=0 \ .
\end{equation}
Hence, we recover the staggered conformal block \eqref{eqn: extended block} by working directly in $d=6$.

\section{The Interacting $\Box^2$ bCFT in $d=6-\epsilon$}
\label{sec: The Interacting Box^2 bCFT in d=6-epsilon}

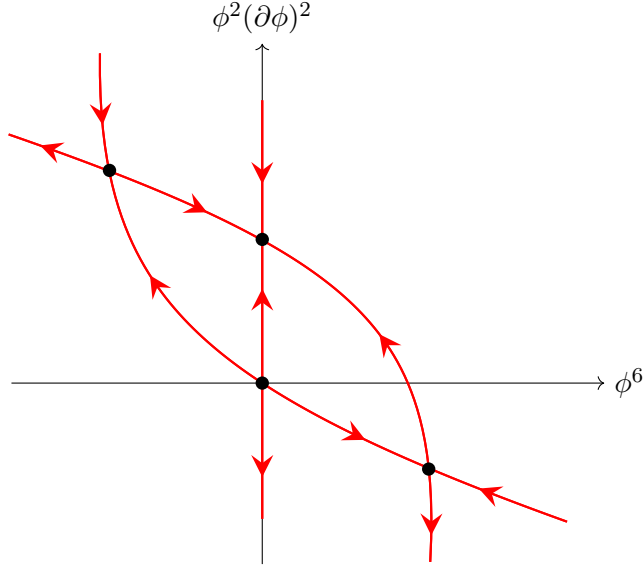
\begin{figure}[ht]
    \centering
    \begin{tikzpicture}[scale=1.5]
        \draw[->] (-2.2, 0) -- (3, 0) node[right] {$\phi^6$};
        \draw[->] (0, -1.6) -- (0, 3) node[above] {$\phi^2(\partial\phi)^2$};

        \draw[red, thick, domain=-2.13:2.9, samples=100, rotate around={-15:(0,0)}] plot (\x, {3/(\x+3)-1}) [arrow inside={end=stealth,opt={red,scale=2}}{0.1,0.7}];
        \draw[red, thick, domain=2.9:-2.13, samples=100, rotate around={-15:(0,0)}] plot (\x, {3/(\x+3)-1}) [arrow inside={end=stealth,opt={red,scale=2}}{0.13,0.68}];
        
        \draw[red, thick, domain=-2.2:2.35, samples=100, rotate around={-15:(0,2)}] plot (\x, {2/(\x-3)+2}) [arrow inside={end=stealth,opt={red,scale=2}}{0.32,0.96}];
        \draw[red, thick, domain=2.35:-2.2, samples=100, rotate around={-15:(0,2)}] plot (\x, {2/(\x-3)+2}) [arrow inside={end=stealth,opt={red,scale=2}}{0.36,0.95}];

        \draw[red, thick] (0, 2.5) -- (0, -1.2) [arrow inside={end=stealth,opt={red,scale=2}}{0.2,0.9}];
        \draw[red, thick] (0, -1.2) -- (0, 2.5) [arrow inside={end=stealth,opt={red,scale=2}}{0.55}];

        \node[circle,fill=black,inner sep=0pt,minimum size=5pt] (Gaussian) at (0,0) {};
        \node[circle,fill=black,inner sep=0pt,minimum size=5pt] (derivative) at (0,1.27) {};
        \node[circle,fill=black,inner sep=0pt,minimum size=5pt] (other 1) at (-1.34,1.88) {};
        \node[circle,fill=black,inner sep=0pt,minimum size=5pt] (other 2) at (1.46,-0.76) {};
    \end{tikzpicture}
    \caption{The space of scalar field theories obtained by deforming the $d=6$ free $\Box^2$ CFT. Four fixed points are present: a Gaussian fixed point, a fixed point with a pure derivative interaction $\phi^2(\partial\phi)^2$, and two fixed points with a mixture of $\phi^2(\partial\phi)^2$ and $\phi^6$ interactions \cite{Safari:2017irw, Safari:2017tgs}.}
    \label{fig: phase diagram of Box^2 theory}
\end{figure}

The space of scalar field theories with a higher derivative kinetic term $\phi\Box^k\phi$ has an interesting phase structure. Consider the critical deformations of the $d$ dimensional free $\Box^k$ CFT. When $k$ and $n-1 = 2k/(d-2k)$ share a common divisor, a pure potential deformation $\phi^{2n}$ cannot be scale-invariant, and a derivative interaction must be included \cite{Safari:2017irw, Safari:2017tgs} (see also refs.\ \cite{dengler1985renormalization,aharony1985novel,aharony1987renormalization}). 
 For example, the $d=6$ free $\Box^2$ CFT has two classically marginal deformations: $\phi^6$ and (schematically) $\phi^2 (\partial \phi)^2$.  
In the epsilon expansion, each of the nontrivial fixed points involve turning on the derivative interaction. The phase diagram from \cite{Safari:2017irw, Safari:2017tgs} is qualitatively illustrated in figure~\ref{fig: phase diagram of Box^2 theory}.

In this work, we will focus on the simplest of these non-trivial fixed points --- the $\Box^2$ CFT with a pure derivative interaction in $d=6-\epsilon$
\begin{equation}
\label{eqn: action again}
S = S_0+\kappa \int \diff^d x\, g_\star \mu^{6-d}\mathcal{O}_\text{int}(x)
\end{equation}
where the derivative interaction
\begin{equation}
    \mathcal{O}_\text{int}\coloneqq\phi^3 \Box \phi + \frac{2}{d-4} \phi^2 (\partial \phi)^2 
\end{equation}
is a primary operator in the $d$ dimensional free $\Box^2$ CFT and $g_\star$ is the critical coupling. 
The $\kappa$ \eqref{eqn: kappa factor} here guarantees that the equation of motion can be expressed in a $\kappa$-free manner.
Note that the presence or absence of a boundary cannot affect the bulk beta function calculation. Therefore, after accounting for the different conventions used in our work and in \cite{Safari:2017irw, Safari:2017tgs}, we can look up the $\epsilon$ expansion of $g_\star$ (see appendix \ref{app: Cross-Check from Conventional Perturbation Theory} for a verification): 
\begin{equation}
\label{eqn: critical coupling}
    g_\star=-\frac{3\epsilon}{8}+O(\epsilon^2)\ . 
\end{equation}

The interacting $\Box^2$ bCFT is defined by the action \eqref{eqn: action again} on $\mathbb{R}^d_+$, with one of the following conformal boundary conditions imposed:
\begin{align}
    \label{eqn: interacting conformal boundary conditions}
    \begin{split}
        &DD:\quad [\Phi^{(0)}]=[\Phi^{(1)}]=0\\
        &DN:\quad [\Phi^{(0)}]=[\Phi^{(2)}]=0\\
        &ND:\quad [\Phi^{(1)}]=[\Phi^{(3)}]=0\\
        &NN:\quad [\Phi^{(2)}]=[\Phi^{(3)}]=0
    \end{split}
\end{align} 
Here, $[\Phi^{(i)}]$ denotes the renormalized boundary primary in the interacting theory, which approaches $\Phi^{(i)}$ in the $\epsilon\to 0$ limit.

We will compute the two-point correlator
\begin{equation}
\label{eqn: phi-phi two-point correlator}
    \braket{\phi(x)\phi(y)}=\frac{f(\xi)}{(4x_ny_n)^{\Delta_\phi}}
\end{equation}
and determine some bCFT data to leading order in their $\epsilon$ expansions. This can, in principle, be done by solving the crossing equation perturbatively in $\epsilon$. By substituting the ansätze \eqref{eqn: general ansatz for bCFT data} for the bCFT data into the crossing equation \eqref{eqn: definition of crossing equation} and expanding both sides around $\epsilon=0$, these bCFT data are constrained by crossing symmetry order by order in $\epsilon$. However, this procedure is not simple to implement in practice because, in the interacting bCFT, an infinite number of primaries usually appear in the OPE and BOE. Consequently, both sides of the crossing equation will contain an infinite number of conformal blocks, requiring a systematic way to analyze the contribution from each conformal family to the bulk/boundary channel.

Another complication is that crossing symmetry alone cannot distinguish the interacting bCFT \eqref{eqn: action again} we study from the other three fixed points shown in figure~\ref{fig: phase diagram of Box^2 theory}. Therefore, we will use the equation of motion
\begin{equation}
    \label{eqn: interacting EOM}
    \Box^2\phi+g_\star\frac{3d-14}{d-4}\left(\phi^2\Box\phi+\frac{1}{3}\Box\phi^3\right)=0
\end{equation}
to facilitate the computation. The equation of motion leads to a differential equation for the undetermined function $f(\xi)$ in \eqref{eqn: phi-phi two-point correlator}. By integrating this differential equation, we obtain the general solution $f(\xi)$, which includes some undetermined integration constants. These constants are fixed by the conformal boundary condition, imposed by matching $f(\xi)$ with its boundary channel decomposition. Through this procedure, we obtain $\braket{\phi\phi}$ and some boundary bCFT data. The bulk bCFT data can then be extracted by matching the now constrained $f(\xi)$ with its bulk channel decomposition.

\subsection{Summary of the Perturbative Results}
\label{subsec: Summary of the Perturbative Results}

While various OPE and BOE coefficients can be found in the text, we tabulate the bulk anomalous dimensions 
\begin{align}
    \Delta_\phi&=1-\frac{\epsilon}{2}+\frac{9\epsilon^2}{1024}+O(\epsilon^3)\label{eqn: anomalous dimension of phi} \ , \\
    \Delta_{\phi^2}&=2-\frac{5\epsilon}{8}+\delta\Delta_{\phi^2}^{(2)}\epsilon^2+O(\epsilon^3)\label{eqn: anomalous dimension of phi^2} \ , \\
    \Delta_{\partial^2\phi^2}&=4-\frac{5\epsilon}{8}+\left(\delta\Delta_{\phi^2}^{(2)}+\frac{1}{64}\right)\epsilon^2+O(\epsilon^3) \ ,
    \label{eqn: anomalous dimension of (d phi)^2}
\end{align}
and the boundary anomalous dimensions in table~\ref{tab: Boundary anomalous dimensions for the four conformal boundary conditions.}.
\begin{table}[h]
\begin{center}
{\renewcommand{\arraystretch}{1.5} 
\begin{tabular}{ |c | c | c | c | c | }
\hline 
 & $DD$ & $DN$ & $ND$ & $NN$ \\ \hline 
$\delta\hat{\Delta}_0^{(1)}$ & - & - & $-\frac{1}{8}$ & $-\frac{1}{16}$ \\
$\delta\hat{\Delta}_1^{(1)}$ & - & $0$ & - & $-\frac{9}{16}$ \\
$\delta\hat{\Delta}_2^{(1)}$ & $-\frac{9}{16}$ & - & $0$ & - \\
$\delta\hat{\Delta}_3^{(1)}$ & $-\frac{1}{16}$ & $-\frac{1}{8}$ & - & - \\
\hline  
\end{tabular}
}
\end{center}
\caption{The $O(\epsilon)$ boundary anomalous dimensions for the four conformal boundary conditions. The subscripts $0,1,2,3$ label the boundary primaries $[\Phi^{(i)}]$'s.}
\label{tab: Boundary anomalous dimensions for the four conformal boundary conditions.}
\end{table}

Note the curious symmetry in the boundary anomalous dimensions.
For the $NN$ and $DD$ cases, we find that $\delta \hat \Delta_i^{(1)} = \delta \hat \Delta_{3-i}^{(1)}$.  The same relation holds for the $ND$ and $DN$ cases.  
For $\phi^4$ theory in $4-\epsilon$ dimensions, a similar relation $\delta \hat \Delta_0^{(1)} = \delta \hat \Delta_{1}^{(1)}$ holds for the anomalous dimensions of
$\hat \phi$ and $\partial_n \hat \phi$, having applied either $N$ or $D$ boundary conditions.  The relation ceases to hold however at $O(\epsilon^2)$
\cite{reeve1981renormalisation,reeve1980critical,diehl1981field,Bissi:2018mcq}.
Looking in more detail at the calculation that leads to this result, one can trace the symmetry to the following facts.  The first is that boundary operators $\Phi^{(i)}$ and $\Phi^{(3-i)}$ are shadow dual and their dimension must add to $d-1$ in the free theory.  The next is that the anomalous dimensions are linearly proportional to the coefficients that multiply the $(1+\xi)^{-1}$ and $(1+\xi)^{-2}$ terms in the free Green's function (see footnote \ref{symmetryfootnote}).  Finally, in moving from $DD$ to $NN$ or from $ND$ to $DN$ boundary conditions, these coefficients flip sign. 
Thus if $\Phi^{(i)}$ has anomalous dimension $\gamma$, then its shadow dual $\Phi^{(3-i)}$, if it existed, must have anomalous dimension $-\gamma$.  However,
in the dual theory it does exist and because the coefficients
in the free Green's function flip sign, its new anomalous dimension will again be $\gamma$.

A way of arguing why the coefficients of $(1+\xi)^{-1}$ and $(1+\xi)^{-2}$ flip sign in flipping both boundary conditions is that the sum of the $DD$ and $NN$ theory (or equivalently the $ND$ and $DN$ theory) is the theory without an interface, for which both of these terms in the Green's function vanish.

To $O(\epsilon)$, our bulk anomalous dimensions \eqref{eqn: anomalous dimension of phi}--\eqref{eqn: anomalous dimension of (d phi)^2} agree with those obtained in \cite{Safari:2017irw, Safari:2017tgs}\footnote{In \cite{Safari:2017irw, Safari:2017tgs}, the $O(\epsilon)$ anomalous dimension of $(\partial\phi)^2$ was computed and shown to match that of $\phi^2$. Since our primary $\partial^2\phi^2$ is a linear combination of $(\partial\phi)^2$ and the descendant $\Box\phi^2$ of $\phi^2$ (see \eqref{eqn: renormalized d^2 phi^2}), their results imply that the $O(\epsilon)$ anomalous dimensions of $\partial^2\phi^2$ and $\phi^2$ should also coincide, as reflected in our \eqref{eqn: anomalous dimension of phi^2} and \eqref{eqn: anomalous dimension of (d phi)^2}.}. We address the subtleties of renormalizing $\partial^2\phi^2$ (and $\Box\phi^2$) in section~\ref{subsec: A Puzzle with partial^2phi^2}. Moreover, our $O(\epsilon)$ analysis of the crossing equation reveals a relationship between the $O(\epsilon^2)$ anomalous dimensions of $\phi^2$ and $\partial^2\phi^2$. Notably, although $\partial^2\phi^2$ is no longer a descendant of $\phi^2$ away from $d=6$, their anomalous dimensions remain the same at $O(\epsilon)$ and only begin to differ from $O(\epsilon^2)$.

\subsection{Using the Equation of Motion}
\label{subsec: Using the Equation of Motion}

By applying the equation of motion \eqref{eqn: interacting EOM} to the $\braket{\phi\phi}$ correlator \eqref{eqn: phi-phi two-point correlator}, we obtain
\begin{equation}
\label{eqn: EOM trick}
    \Box_x^2\left(\frac{f(\xi)}{(4x_ny_n)^{\Delta_\phi}}\right)=-g_\star\frac{3d-14}{d-4}\braket{\left(\phi^2\Box\phi+\frac{1}{3}\Box\phi^3\right)(x)\phi(y)}
\end{equation}
This differential equation can be solved perturbatively in $\epsilon$. By substituting the ans\"atze
\begin{equation}
    f(\xi)=f_0(\xi)+g_\star f_1(\xi)+O(g_\star^2),\quad \Delta_\phi=\Delta_\phi^{(0)}+g_\star  \Delta_\phi^{(1)}+O(g_\star^2)
\end{equation}
the left-hand side of \eqref{eqn: EOM trick} can be expanded in powers of $g_\star$ 
\begin{equation}
\label{eqn: expansion of EOM trick in g}
    \Box_x^2\left(\frac{f_0(\xi)}{(4x_ny_n)^{\Delta_\phi^{(0)}}}\right)+g_\star\Box_x^2\left(\frac{f_1(\xi)-\Delta_\phi^{(1)}\log(4x_ny_n)f_0(\xi)}{(4x_ny_n)^{\Delta_\phi^{(0)}}}\right)+O(g_\star^2) \ .
\end{equation}
The first term in \eqref{eqn: expansion of EOM trick in g} vanishes by the free equation of motion $\Box^2\phi=0$. By substituting $d=6-\epsilon$ and the $\epsilon$ expansion of $g_\star$, \eqref{eqn: expansion of EOM trick in g} can be expanded in powers of $\epsilon$
\begin{equation}
\label{eqn: O(epsilon) LHS of EOM trick}
    -\frac{3\epsilon}{8}\left[\left.\Box_x^2\left(\frac{f_1(\xi)}{(4x_ny_n)^{\Delta_\phi^{(0)}}}\right)\right\rvert_{d=6}-\Delta_\phi^{(1)}\left.\Box_x^2\left(\frac{\log(4x_ny_n)f_0(\xi)}{(4x_ny_n)^{\Delta_\phi^{(0)}}}\right)\right\rvert_{d=6}\right]+O(\epsilon^2)
\end{equation}
where
\begin{equation}
\label{eqn: 1-loop EOM equation LHS 1}
\begin{split}
    \left.\Box_x^2\left(\frac{f_1(\xi)}{(4x_ny_n)^{\Delta_\phi^{(0)}}}\right)\right\rvert_{d=6}&=\frac{1}{4x_n^5y_n}\biggl[\xi^2(1+\xi)^2f^{(4)}_1(\xi)+8\xi(1+\xi)(1+2\xi)f_1^{(3)}(\xi)\\
    &+12(1+6\xi+6\xi^2)f_1''(\xi)+48(1+2\xi)f_1'(\xi)+24f_1(\xi)\biggl]
\end{split}
\end{equation}
and
\begin{equation}
    \label{eqn: 1-loop EOM equation LHS 2}
    \begin{split}
        \left.\Box_x^2\left(\frac{\log(4x_ny_n)f_0(\xi)}{(4x_ny_n)^{\Delta_\phi^{(0)}}}\right)\right\rvert_{d=6}
    &=\frac{1}{4x_n^5y_n}\biggl[2 \xi  (1+\xi) \left(\frac{x_n}{y_n}-(1+2\xi)\right) \left.f_0^{(3)}(\xi )\right\rvert_{d=6} \\
    &+ \left(\frac{x_n^2}{y_n^2}+6 (1+2 \xi)
   \frac{x_n}{y_n}-7 (1+6\xi+6\xi^2)\right) \left.f_0''(\xi )\right\rvert_{d=6}\\
   &+4\left(3 \frac{x_n}{y_n}-13 (1+2\xi)\right)\left.f_0'(\xi )\right\rvert_{d=6}-50 \left.f_0(\xi )\right\rvert_{d=6}\biggl]
    \ .
    \end{split}
\end{equation}

The two-point correlator on the right-hand side of \eqref{eqn: EOM trick} can be computed order by order in $g_\star$ using diagrammatic perturbation theory. Substituting $d=6-\epsilon$ and the $\epsilon$ expansion of $g_\star$, the right-hand side of \eqref{eqn: EOM trick} can be expanded in powers of $\epsilon$
\begin{equation}
\label{eqn: O(epsilon) RHS of EOM trick}
    -\left(-\frac{3\epsilon}{8}\right)\cdot2\cdot\left.\left(\braket{\phi^2(x)\Box_x\phi(x)\phi(y)}_0+\frac{1}{3}\braket{\Box_x\phi^3(x)\phi(y)}_0\right)\right\rvert_{d=6}+O(\epsilon^2) \ .
\end{equation}
The free two-point correlator has the general form
\begin{equation}
\label{eqn: two-point correlator on the RHS of EOM trick}
    \braket{\phi^2(x)\Box_x\phi(x)\phi(y)}_0+\frac{1}{3}\braket{\Box_x\phi^3(x)\phi(y)}_0=\frac{H(\xi,x_n/y_n)}{(2x_n)^{3\Delta_\phi^{(0)}+2}(2y_n)^{\Delta_\phi^{(0)}}}
\end{equation}
dictated by the Ward identities for $P_a$, $M_{ab}$, and $D$. $H(\xi,x_n/y_n)$ can be computed by Wick contractions. In $d=6$, \eqref{eqn: two-point correlator on the RHS of EOM trick} has the same kinematic pre-factor, $x_n^{-5}y_n^{-1}$, as \eqref{eqn: 1-loop EOM equation LHS 1} and \eqref{eqn: 1-loop EOM equation LHS 2}, which can be factored out when matching the left-hand side \eqref{eqn: O(epsilon) LHS of EOM trick} and right-hand side \eqref{eqn: O(epsilon) RHS of EOM trick} of \eqref{eqn: EOM trick} at $O(\epsilon)$.

Through the $O(\epsilon)$ matching, we can extract the $O(g_\star)$ anomalous dimension $\Delta_\phi^{(1)}$ and derive a differential equation for $f_1(\xi)$. More explicitly, $\Delta_\phi^{(1)}$ is determined by matching the $x_n/y_n$ dependences of \eqref{eqn: 1-loop EOM equation LHS 2} and \eqref{eqn: two-point correlator on the RHS of EOM trick}. In general, this requires us to specialize to a conformal boundary condition. In other words, we need to pick a $f_0(\xi)$ from \eqref{eqn: correlator solution to the crossing equation} to proceed. However, in our case, we can make further progress before focusing on a specific conformal boundary condition. Note that the operator
\begin{equation}
\label{eqn: bulk primary phi (d phi)^2}
    \phi^2\Box\phi+\frac{1}{3(d-5)}\Box\phi^3
\end{equation}
is a bulk primary in the free $\Box^2$ bCFT. Thus, \eqref{eqn: two-point correlator on the RHS of EOM trick} is a correlator of two $d=6$ primaries. The Ward identity for $K_a$ further imposes that $H(\xi)$ must be independent of $x_n/y_n$. This observation allows us to conclude that 
\begin{equation}
\label{eqn: one-loop anomalous dimension of phi}
    \Delta_\phi^{(1)}=0 \ .
\end{equation}

\subsection{Using the Boundary Channel}
\label{subsec: Using the Boundary Channel}

For the remainder of section~\ref{sec: The Interacting Box^2 bCFT in d=6-epsilon}, we will focus on the $NN$ boundary condition. A parallel analysis can be done for the $DD$, $DN$, and $ND$ boundary conditions without difficulty.

Recall that the $\braket{\phi\phi}_0$ correlator with the $NN$ boundary condition imposed is
\begin{equation}
    \braket{\phi(x)\phi(y)}_0=\frac{f_0(\xi)}{(4x_ny_n)^{\Delta_\phi^{(0)}}}=\frac{1}{(4x_ny_n)^\frac{d-4}{2}}\left(\xi^{-\frac{d-4}{2}}+\frac{d-4}{2}(1+\xi)^{-\frac{d-2}{2}}+(1+\xi)^{-\frac{d-4}{2}}\right) \ .
\end{equation}
By Wick contractions, the $d=6$ free two-point correlator on the right-hand side of \eqref{eqn: EOM trick} is
\begin{equation}
    \left.\left(\braket{\phi^2(x)\Box_x\phi(x)\phi(y)}_0+\frac{1}{3}\braket{\Box_x\phi^3(x)\phi(y)}_0\right)\right\rvert_{d=6}=\frac{1}{(2x_n)^52y_n}\frac{8(-2-3\xi+2\xi^3)}{\xi^2(1+\xi)^2} \ .
\end{equation}
After imposing \eqref{eqn: one-loop anomalous dimension of phi}, the differential equation \eqref{eqn: EOM trick} at $O(\epsilon)$ is
\begin{equation}
\label{eqn: ODE for f1}
\begin{split}
    &\xi^2(1+\xi)^2f^{(4)}_1+8\xi(1+\xi)(1+2\xi)f_1^{(3)}+12(1+6\xi+6\xi^2)f_1''+48(1+2\xi)f_1'+24f_1\\
   &=\frac{2+3\xi-2\xi^3}{\xi^2(1+\xi)^2}
\end{split}
\end{equation}
which can be integrated to obtain
\begin{equation}
\label{eqn: general solution for f_1}
    \begin{split}
        f_1(\xi)&=\frac{c_1}{\xi}+\frac{c_2}{\xi^2}+\frac{c_3}{1+\xi}+\frac{c_4}{(1+\xi)^2}\\
        &+\frac{1}{18 \xi ^2 (1+\xi)^2}\biggr(11 \xi ^3-3 \xi ^3 \log (\xi )-3 \xi ^3 \log (1+\xi)+21 \xi ^2-18 \xi ^2 \log (\xi)\\
        &-18 \xi ^2 \log (1+\xi)+48 \xi -27 \xi 
        \log (1+\xi)-12 \log (1+\xi)+14\biggr) \ .
    \end{split}
\end{equation}
In this section, we fix the integration constants $c_i$'s by imposing the $NN$ boundary condition on \eqref{eqn: general solution for f_1}.

Imposing a conformal boundary condition is equivalent to removing the contributions of the corresponding boundary primaries to the boundary channel
\begin{equation}
\label{eqn: boundary channel decomposition of f}
    f(\xi)=\sum_\alpha\mu_{\alpha}^2 G_\text{bry}(\hat{\Delta}_\alpha;\xi) \ .
\end{equation}
We have already determined the left-hand side of \eqref{eqn: boundary channel decomposition of f}
\begin{equation}
\label{eqn: EOM decomposition of f}
    f(\xi)=f_0(\xi)+g_\star f_1(\xi)+O(g_\star^2)
\end{equation}
to $O(\epsilon)$ in terms of four integration constants $c_i$'s. By substituting the ans\"atze
\begin{equation}
        \mu_\alpha^2=(\mu_\alpha^2)^{(0)}+\delta(\mu_\alpha^2)^{(1)}\epsilon+O(\epsilon^2),\quad \hat{\Delta}_\alpha=\hat{\Delta}_\alpha^{(0)}+\delta\hat{\Delta}_\alpha^{(1)}\epsilon+O(\epsilon^2)
\end{equation}
the right-hand side of \eqref{eqn: boundary channel decomposition of f} can also be expanded to $O(\epsilon)$ in terms of undetermined $O(\epsilon)$ boundary bCFT data $\{\delta(\mu_\alpha^2)^{(1)},\delta\hat{\Delta}_\alpha^{(1)}\}$. By matching the two $\epsilon$ expansions, we can simultaneously impose the conformal boundary condition and extract $\{\delta(\mu_\alpha^2)^{(1)},\delta\hat{\Delta}_\alpha^{(1)}\}$.

Since, in general, there are an infinite number of boundary primaries appearing in the interacting BOE of $\phi$, we need a systematic way to deal with the infinite sum in \eqref{eqn: boundary channel decomposition of f}. Our strategy is to expand boundary conformal blocks in \eqref{eqn: boundary channel decomposition of f} around $\xi=\infty$ 
\begin{equation}
    G_\text{bry}(\hat{\Delta}_\alpha;\xi)=\xi^{-\hat{\Delta}_\alpha}\left(1+O(\xi^{-1})\right) \ .
\end{equation}
Boundary primaries $\hat{\mathcal{O}}_\alpha$ with higher scaling dimensions $\hat{\Delta}_\alpha$ enter the boundary channel $f(\xi)$ at higher orders in $\xi^{-1}$. In this way, we can truncate the double expansion of \eqref{eqn: boundary channel decomposition of f} at $O(\epsilon,\xi^{-n})$ for some finite $n\in\mathbb{N}$ and study the contributions of a finite number of boundary primaries to the $O(\epsilon)$ boundary channel. This double expansion should be matched with the corresponding double expansion of \eqref{eqn: EOM decomposition of f} to solve for the $c_i$'s and $\{\delta(\mu_\alpha^2)^{(1)},\delta\hat{\Delta}_\alpha^{(1)}\}$.

We can make an a priori argument that, to $O(\epsilon)$, only two boundary primaries will contribute to the boundary channel \eqref{eqn: boundary channel decomposition of f}. The argument proceeds as follows. Suppose a boundary primary $\hat{\mathcal{O}}_\alpha$ appears in the interacting BOE of $\phi$  but not in the free BOE. Then the BOE coefficient $\mu_\alpha$ is at least of order $g_\star$, and so the contribution of $\hat{\mathcal{O}}_\alpha$ to the boundary channel is at least of order $g_\star^2\sim\epsilon^2$. Hence, $\hat{\mathcal{O}}_\alpha$ contributes to the $O(\epsilon)$ boundary channel if and only if it appears in the free BOE of $\phi$. This means $\hat{\mathcal{O}}_\alpha$ must be one of the four $[\Phi^{(i)}]$'s. The conformal boundary condition removes two of them from the boundary spectrum, leaving only the remaining two to contribute to the $O(\epsilon)$ boundary channel. Under the $NN$ boundary condition, the two boundary primaries that contribute are $[\Phi^{(0)}]$ and $[\Phi^{(1)}]$.

Guided by the above argument, we match the double expansion of
\begin{equation}
\label{eqn: boundary channel to O(epsilon)}
    \mu_0^2 G_\text{bry}(\hat{\Delta}_0;\xi)+\mu_1^2 G_\text{bry}(\hat{\Delta}_1;\xi)
\end{equation}
with that of \eqref{eqn: EOM decomposition of f} to $O(\epsilon,\xi^{-4})$. The results are\footnote{%
\label{symmetryfootnote}
  To get a better understanding of the peculiar symmetry of table \ref{tab: Boundary anomalous dimensions for the four conformal boundary conditions.}, it is useful to redo the calculation for a more general version of the tree level $\langle \phi \phi \rangle$ two-point function that includes the four conformal choices as special cases.  If we replace (\ref{eqn: correlator solution to the crossing equation}) with 
  \[
  f_0(\xi) = \xi^{-\frac{d-4}{2}} + \alpha (1+\xi)^{-\frac{d-2}{2}} + \beta (1+\xi)^{-\frac{d-4}{2}} \ ,
  \]
  then the leading order anomalous dimensions for the boundary operators take the form
  \[
\delta\hat{\Delta}_0^{(1)} = - \delta\hat{\Delta}_3^{(1)} = \frac{1}{2} (\alpha - 2 \beta) \ , \; \; \;
\delta\hat{\Delta}_1^{(1)} = - \delta\hat{\Delta}_2^{(1)} = - \frac{9 \alpha}{16} \ .
  \]
  One sees that $\Phi^{(i)}$ and $\Phi^{(3-i)}$ have anomalous dimensions that sum to zero at leading order in $\epsilon$, consistent with being shadow dual.
}
\begin{equation}
\label{eqn: results of matching boundary channel and EOM decomposition}
    \begin{aligned}
        \delta\hat{\Delta}_0^{(1)}&=-\frac{1}{16},   &\quad   
        c_1&=  \frac{7}{18} - \frac{4}{3} \delta(\mu_0^2)^{(1)} ,
        &\quad   
        c_2&= \frac{-61+24\delta(\mu_0^2)^{(1)}-48\delta(\mu_1^2)^{(1)}}{36},
        \\
        \delta\hat{\Delta}_1^{(1)}&=-\frac{9}{16},   &\quad   
        c_3&= -1 - \frac{4}{3}  \delta(\mu_0^2)^{(1)} ,   
        &\quad   
        c_4&= \frac{9 - 8  \delta(\mu_0^2)^{(1)} - 16  \delta(\mu_1^2)^{(1)} }{12} \ .
\end{aligned}
\end{equation}
After imposing \eqref{eqn: results of matching boundary channel and EOM decomposition}, the $\epsilon$ expansion of \eqref{eqn: boundary channel to O(epsilon)} is verified to match that of \eqref{eqn: EOM decomposition of f} to $O(\epsilon)$, thereby providing a posteriori justification for our argument.

We are left with two undetermined BOE coefficients, $\delta(\mu_0^2)^{(1)}$ and $\delta(\mu_1^2)^{(1)}$. These coefficients can be determined by looking at the $\xi\to 0$ limit of $f(\xi)$, where it should behave as
\begin{equation}
    f(\xi)=\xi^{-\Delta_\phi}+\text{higher-order terms in $\xi$} \ .
\end{equation}
The leading-order term corresponds to the bulk identity conformal block. We do not yet know the higher-order terms, as they involve bulk bCFT data that will be computed in section~\ref{subsec: Reconstructing the Bulk Channel}. However, by matching the $\epsilon$ expansion of this leading-order term $\xi^{-\Delta_\phi}$ with the double expansion (around $\epsilon=0$ and $\xi=0$) of \eqref{eqn: boundary channel to O(epsilon)} to $O(\epsilon)$, we can solve for the two BOE coefficients. The results, along with those for the other three conformal boundary conditions, are summarized in table~\ref{tab: The O(epsilon) BOE coefficients for the four conformal boundary conditions.}.

\begin{table}[h]
\begin{center}
{\renewcommand{\arraystretch}{1.5} 
\begin{tabular}{ |c | c | c | c | c | }
\hline 
 & $DD$ & $DN$ & $ND$ & $NN$ \\ \hline 
$\delta(\mu_0^2)^{(1)}$ & - & - & $\frac{1}{8}$ & $\frac{5}{8}$ \\
$\delta(\mu_1^2)^{(1)}$ & - & $\frac{3}{16}$ & - & $-\frac{3}{8}$ \\
$\delta(\mu_2^2)^{(1)}$ & $\frac{3}{32}$ & - & $-\frac{3}{16}$ & - \\
$\delta(\mu_3^2)^{(1)}$ & $\frac{13}{192}$ & $-\frac{7}{48}$ & - & - \\
\hline  
\end{tabular}
}
\end{center}
\caption{The $O(\epsilon)$ BOE coefficients for the four conformal boundary conditions.}
\label{tab: The O(epsilon) BOE coefficients for the four conformal boundary conditions.}
\end{table}
Some of the $\delta (\mu^2_\alpha)^{(1)}$'s are negative. Note that it is fundamentally the $\mu_\alpha$'s that receive perturbative corrections: $\mu_\alpha=\mu_\alpha^{(0)}+\delta\mu_\alpha^{(1)}\epsilon+O(\epsilon^2)$. Thus, $\delta(\mu^2_\alpha)^{(1)}=2\mu_\alpha^{(0)}\delta\mu_\alpha^{(1)}$ is not a squared term.

Under the $NN$ boundary condition, the $\epsilon$ expansion of $f(\xi)$ in the $\braket{\phi\phi}$ correlator \eqref{eqn: phi-phi two-point correlator} is now fixed to $O(\epsilon)$: 
\begin{equation}
\label{eqn: solution}
    \begin{split}
        f(\xi)&=\frac{1}{\xi }+\frac{1}{(1+\xi)^2}+\frac{1}{1+\xi}\\
        &+\frac{\epsilon}{16 \xi
   ^2 (1+\xi)^2}\biggr(  10 \xi ^3+9
   \xi ^3 \log (\xi )+9 \xi ^3 \log (1+\xi)+\xi ^2+22 \xi ^2 \log (\xi )\\
   &+22 \xi ^2 \log
   (1+\xi)-4 \xi +8 \xi  \log (\xi )+9 \xi  \log (1+\xi)+4 \log (1+\xi)\biggr)+O(\epsilon^2)
   \ .
    \end{split}
\end{equation}

\subsection{Reconstructing the Bulk Channel}
\label{subsec: Reconstructing the Bulk Channel}

In this section, we propose the ans\"atze for the $\epsilon$ expansions of the bulk bCFT data and determine some of this data  by matching \eqref{eqn: solution} with the $\epsilon$ expansion of
\begin{equation}
\label{eqn: bulk channel decomposition of f (d neq 6)}
f(\xi)=\sum_k\lambda a_k G_\text{bulk}(\Delta_k;\xi).
\end{equation}
In general, an infinite number of bulk primaries appear in the interacting OPE $\phi \times \phi$. To handle the infinite sum in \eqref{eqn: bulk channel decomposition of f (d neq 6)}, we expand the bulk conformal blocks around $\xi = 0$
\begin{equation}
G_\text{bulk}(\Delta_k;\xi)=\xi^{\frac{1}{2}(\Delta_k-2\Delta_\phi)}\left(1+O(\xi)\right).
\end{equation}
Bulk primaries $\mathcal{O}_k$ with higher scaling dimensions $\Delta_k$ enter the bulk channel $f(\xi)$ at higher orders in $\xi$. In this way, we can truncate the double expansion of \eqref{eqn: bulk channel decomposition of f (d neq 6)} at $O(\epsilon, \xi^n)$ for some finite $n$ and study the contributions of a finite number of bulk primaries to the $O(\epsilon)$ bulk channel.

To motivate our ans\"atze, we study the $\epsilon\to 0$ limit of the bulk channel decomposition \eqref{eqn: bulk channel decomposition of f (d neq 6)}. We first establish the notation:
\begin{equation}
\label{eqn: explicit bulk channel decomposition in d=6-epsilon}
    f(\xi)\simeq G_\text{bulk}(0;\xi)+\lambda a_{\phi^2}G_\text{bulk}(\Delta_{\phi^2};\xi)+\lambda a_{\partial^2\phi^2}G_\text{bulk}(\Delta_{\partial^2\phi^2};\xi)+\sum_{n=3}^\infty\lambda a_{2n} G_\text{bulk}(2n;\xi)
\end{equation}
where $\simeq$ means the equation holds to $O(\epsilon)$. The first three terms in \eqref{eqn: explicit bulk channel decomposition in d=6-epsilon} represent the contributions from $I$, $\phi^2$, and $\partial^2\phi^2$, respectively. These are the bulk primaries that appear in the free $\phi \times \phi$ OPE. 

Note that while we write $\lambda a_{\partial^2 \phi^2}$, at this order in $\epsilon$ we cannot separate
small contributions from $\phi^4$, which has the same leading-order scaling dimension.
The infinite sum in \eqref{eqn: explicit bulk channel decomposition in d=6-epsilon} captures the further contributions from those bulk primaries that appear in the interacting but not the free $\phi \times \phi$ OPE. We classify their contributions according to their scaling dimensions in the $d=6$ free theory. This classification is possible because the $O(\epsilon)$ bulk channel cannot distinguish their conformal blocks. For example, $\phi^6$ and $\mathcal{O}_\text{int}$ are two bulk primaries that can appear in the interacting $\phi \times \phi$ OPE, and both have $d=6$ scaling dimension of $6$. The fact that they do not appear in the free $\phi \times \phi$ OPE implies both $\lambda a_{\phi^6}$ and $\lambda a_{\mathcal{O}_\text{int}}$ are at least of order $g_\star \sim \epsilon$. Therefore, their anomalous dimensions do not enter the $O(\epsilon)$ bulk channel
\begin{equation}
    f(\xi)\supset \lambda a_{\phi^6}G(\Delta_{\phi^6};\xi)+\lambda a_{\mathcal{O}_\text{int}}G(\Delta_{\mathcal{O}_\text{int}};\xi)\simeq (\delta \lambda a_{\phi^6}^{(1)}+\delta \lambda a_{\mathcal{O}_\text{int}}^{(1)})\epsilon\cdot G_\text{bulk}(6;\xi) \ .
\end{equation}
In fact, they are the only two bulk primaries with $d=6$ scaling dimension of $6$ that can have a non-zero $\phi\times\phi$ OPE coefficient.\footnote{%
For a bulk primary to contribute to the bulk channel, it must be a scalar. Candidates for bulk scalar primaries with $d=6$ scaling dimensions of $6$ must take one of three forms: $\phi^6$, $\partial^2\phi^4$, or $\partial^4\phi^2$. We can identify $\partial^2\phi^4$ as $\mathcal{O}_\text{int}$. There are no bulk scalar primaries bilinear in $\phi$ with more than two derivatives \cite{Brust:2016gjy}.}
Hence, we propose the ansatz
\begin{equation}
    \lambda a_6=\delta\lambda a_6^{(1)}\epsilon+O(\epsilon^2)
\end{equation}
and identify 
\begin{equation}
    \delta\lambda a_6^{(1)}=\delta \lambda a_{\phi^6}^{(1)}+\delta \lambda a_{\mathcal{O}_\text{int}}^{(1)} \ .
\end{equation}
More generally, a bulk primary contributes to the bulk channel if and only if its $d=6$ scaling dimension is even. This follows from the $\mathbb{Z}_2$ symmetry and the scalar condition. Thus, we propose 
\begin{equation}
\label{eqn: O(epsilon) ansatze of O_2n}
        \lambda a_{2n}=\delta\lambda a_{2n}^{(1)}\epsilon+O(\epsilon^2) \ .
\end{equation}

The left-hand side of \eqref{eqn: explicit bulk channel decomposition in d=6-epsilon}, that is, $f(\xi)$ itself, has a well-defined $\epsilon\to 0$ limit
\begin{equation}
\label{eqn: explicit bulk channel decomposition in d=6}
\lim_{\epsilon\to 0}f(\xi)=\left.f_0(\xi)\right\rvert_{d=6}=\left.G_\text{bulk}(0;\xi)\right\rvert_{d=6}+G_\text{sta}(\xi) \ .
\end{equation}
However, not every individual term on the right-hand side of \eqref{eqn: explicit bulk channel decomposition in d=6-epsilon} does. The infinite sum in \eqref{eqn: explicit bulk channel decomposition in d=6-epsilon} vanishes in the $\epsilon\to 0$ limit due to the $\lambda a_{2n}$ ansatz \eqref{eqn: O(epsilon) ansatze of O_2n}. The bulk identity block has a well-defined $\epsilon\to 0$ limit
\begin{equation}
\lim_{\epsilon\to 0}G_\text{bulk}(0;\xi)=\left.G_\text{bulk}(0;\xi)\right\rvert_{d=6} \ .
\end{equation}
Therefore, by comparing \eqref{eqn: explicit bulk channel decomposition in d=6-epsilon} and \eqref{eqn: explicit bulk channel decomposition in d=6}, we conclude that
\begin{equation}
\label{eqn: interacting theory well-defined limit}
\lim_{\epsilon\to 0}\left[\lambda a_{\phi^2}G_\text{bulk}(\Delta_{\phi^2};\xi)+\lambda a_{\partial^2\phi^2}G_\text{bulk}(\Delta_{\partial^2\phi^2};\xi)\right]=G_\text{sta}(\xi) \ .
\end{equation}

Similar to the situation in the free theory, the limit does not commute with the sum because the $\epsilon\to 0$ limit of each term is not well-defined. The first term in \eqref{eqn: interacting theory well-defined limit} has an $\epsilon^{-1}$ pole due to the $\phi^2$ conformal block. By substituting the ansatz
\begin{equation}
\label{eqn: O(epsilon) ansatze of phi^2}
    \begin{aligned}
        \Delta_\phi&=\Delta_\phi^{(0)}+\delta \Delta_\phi^{(2)}\epsilon^2+O(\epsilon^3) \ ,\\
        \Delta_{\phi^2}&=\Delta_{\phi^2}^{(0)}+\delta \Delta_{\phi^2}^{(1)}\epsilon+\delta \Delta_{\phi^2}^{(2)}\epsilon^2+\delta \Delta_{\phi^2}^{(3)}\epsilon^3+O(\epsilon^4) \ ,\\
        \lambda a_{\phi^2}&=\lambda a_{\phi^2}^{(0)}+\delta \lambda a_{\phi^2}^{(1)}\epsilon+\delta \lambda a_{\phi^2}^{(2)}\epsilon^2+O(\epsilon^3)
        \ ,
    \end{aligned}
\end{equation}
this $\epsilon^{-1}$ pole is
\begin{equation}
\label{eqn: pole of phi^2 block in the interacting theory}
    \lambda a_{\phi^2}G_\text{bulk}(\Delta_{\phi^2};\xi)=\frac{2\lambda a_{\phi^2}^{(0)}}{(1-2\delta \Delta_{\phi^2}^{(1)})\epsilon}\frac{\xi}{(1+\xi)^2}+O(\epsilon^0) \ .
\end{equation}
Note that the residue of the pole is shifted compared to that of the free theory
\begin{equation}
\label{eqn: pole of phi^2 block in the free theory}
    \lambda a_{\phi^2}^{(0)}G_\text{bulk}(\Delta_{\phi^2}^{(0)};\xi)=\frac{2\lambda a_{\phi^2}^{(0)}}{\epsilon}\frac{\xi}{(1+\xi)^2}+O(\epsilon^0)
\end{equation}
due to the $O(\epsilon)$ anomalous dimension $\delta \Delta_{\phi^2}^{(1)}$. To ensure that the limit of the sum is well-defined, the $\epsilon^{-1}$ pole in \eqref{eqn: pole of phi^2 block in the interacting theory} must be canceled by an $\epsilon^{-1}$ pole in the second term of \eqref{eqn: interacting theory well-defined limit}. Since the $\partial^2\phi^2$ conformal block does not contain any $\epsilon^{-1}$ pole, this pole must come from the OPE coefficient $\lambda a_{\partial^2\phi^2}$. The free theory OPE coefficient $\lambda a_{\partial^2\phi^2}^{(0)}\sim \epsilon^{-1}$ can only cancel the $\epsilon^{-1}$ pole in \eqref{eqn: pole of phi^2 block in the free theory}. Hence, we must include an $O(\epsilon^{-1})$ anomalous OPE coefficient $\delta\lambda a_{\partial^2\phi^2}^{(-1)}$ in the ans\"atze
\begin{equation}
\label{eqn: O(epsilon) ansatze of (d phi)^2}
    \begin{aligned}
        \Delta_{\partial^2\phi^2}&=\Delta_{\partial^2\phi^2}^{(0)}+\delta \Delta_{\partial^2\phi^2}^{(1)}\epsilon+\delta \Delta_{\partial^2\phi^2}^{(2)}\epsilon^2+O(\epsilon^3) \ , \\
        \lambda a_{\partial^2\phi^2}&=\lambda a_{\partial^2\phi^2}^{(0)}+\frac{\delta\lambda a_{\partial^2\phi^2}^{(-1)}}{\epsilon}+\delta\lambda a_{\partial^2\phi^2}^{(0)}+\delta\lambda a_{\partial^2\phi^2}^{(1)}\epsilon+O(\epsilon^2) \ .
    \end{aligned}
\end{equation}
In this notation, $\Delta^{(0)}_{\partial^2 \phi^2}$ and $\lambda a^{(0)}_{\partial^2 \phi^2}$ take their tree-level values in $d$ dimensions.  The
quantities $\delta \Delta^{(i)}_{\partial^2 \phi^2}$ and 
$\delta \lambda a^{(i)}_{\partial^2 \phi^2}$ are $O(1)$ coefficients of an $\epsilon$ expansion that can in principle be deduced from a perturbative computation in $g$.  
Note that the $\epsilon^{-1}$ pole in 
\begin{equation}
    \lambda a_{\partial^2\phi^2} G_\text{bulk}(\Delta_{\partial^2\phi^2};\xi)=\left(\lambda a_{\partial^2\phi^2}^{(0)}+\frac{\delta\lambda a_{\partial^2\phi^2}^{(-1)}}{\epsilon}\right)\frac{\xi}{(1+\xi)^2}+O(\epsilon^0)
\end{equation}
cancels the $\epsilon^{-1}$ pole in \eqref{eqn: pole of phi^2 block in the interacting theory}:
\begin{equation}
\label{eqn: condition for a smooth epsilon to 0 limit}
    \lim_{\epsilon\to 0}\left(\frac{2\lambda a_{\phi^2}^{(0)}}{1-2\delta \Delta_{\phi^2}^{(1)}}+\lambda a_{\partial^2\phi^2}^{(0)}\epsilon+\delta\lambda a_{\partial^2\phi^2}^{(-1)}\right)=0 \ .
\end{equation}
This condition determines the $O(\epsilon^{-1})$ OPE coefficient $\delta\lambda a_{\partial^2\phi^2}^{(-1)}$ in terms of the $O(\epsilon)$ anomalous dimension $\delta \Delta_{\phi^2}^{(1)}$. We will address the origin of this peculiar $O(\epsilon^{-1})$ term from the perspective of conventional perturbation theory in section~\ref{subsec: A Puzzle with partial^2phi^2}. For future reference, we rewrite \eqref{eqn: condition for a smooth epsilon to 0 limit} as
\begin{equation}
\label{eqn: ratio of lambda a for partial^2 phi^2}
    \lim_{\epsilon\to 0}\frac{\lambda a_{\partial^2\phi^2}^{(0)}}{\lambda a_{\partial^2\phi^2}^{(0)}+\delta\lambda a_{\partial^2\phi^2}^{(-1)}\epsilon^{-1}}=1-2\delta\Delta_{\phi^2}^{(1)}
\end{equation}
by noting the relation
\begin{equation}
    \lim_{\epsilon\to 0}2\lambda a_{\phi^2}^{(0)}=-\lim_{\epsilon\to 0}\lambda a_{\partial^2\phi^2}^{(0)}\epsilon
\end{equation}
which directly follows from \eqref{eqn: residue of the phi^2 block in the free theory}.

The staggered block that appears in the $d\to 6$ limit imposes two generic constraints on the CFT data.  Taking both a small $\epsilon$ and small $\xi$ expansion of $f(\xi)$, there will be terms that are order zero in the small $\epsilon$ expansion and of order $\log \xi$ in the small $\xi$ expansion.  For these terms to vanish,
it must be that $\delta \Delta_{\phi^2}^{(1)} = \delta \Delta_{\partial^2 \phi^2}^{(1)}$.  
Vanishing of terms that are both order zero in $\epsilon$ and order zero in $\xi$ then fixes the order zero in $\epsilon$ part of $\delta \lambda a_{\partial^2 \phi^2}^{(0)}$ in terms of 
$\delta \lambda  a_{ \phi^2}^{(1)}$, 
$\delta \Delta_{\phi^2}^{(1)}$, and
$\delta \Delta_{\phi^2}^{(2)}$.

By substituting the ans\"atze \eqref{eqn: O(epsilon) ansatze of O_2n}, \eqref{eqn: O(epsilon) ansatze of phi^2}, and \eqref{eqn: O(epsilon) ansatze of (d phi)^2} into \eqref{eqn: explicit bulk channel decomposition in d=6-epsilon} and expanding to $O(\epsilon)$, we obtain the bulk channel decomposition of $f(\xi)$ to $O(\epsilon)$. This expansion can then be matched with \eqref{eqn: solution} to determine some of the bulk bCFT data under the $NN$ boundary condition. The results include
\begin{align}  
    &\delta \Delta_{\phi^2}^{(1)}=\delta \Delta_{\partial^2\phi^2}^{(1)}=\frac{3}{8} \label{eqn: O(epsilon) anomalous dimensions relation} \ , \\
    &\delta\lambda a_{\phi^2}^{(1)}=1,\quad \delta\lambda a_{\partial^2\phi^2}^{(-1)}=-12  \ ,
    \label{eqn: OPE coeffs}
\end{align}
as well as an infinite number of OPE coefficients $\delta\lambda a_{2n}^{(1)}$ for $n\geq 3$. Since these OPE coefficients $\delta\lambda a_{2n}^{(1)}$ generally receive contributions from several bulk primaries with the same $d=6$ scaling dimension, they are not useful and will not be tabulated. However, one of them, 
\begin{equation}
\label{eqn: NN result of the O(epsilon) OPE of 6}
    \delta\lambda a_6^{(1)}=0 \ ,
\end{equation}
can be straightforwardly verified using conventional perturbation theory. We elaborate on this cross-check in appendix~\ref{app: Cross-Check from Conventional Perturbation Theory}.

In addition to the above results, we have derived several constraints on the bulk bCFT data, two of which are particularly noteworthy:
\begin{align}  
    \delta\Delta_{\partial^2\phi^2}^{(2)}&=\delta\Delta_{\phi^2}^{(2)}+\frac{1}{64} \label{eqn: constraint 1 from the O(epsilon) analysis} \ ,\\
    \delta\lambda a_{\partial^2\phi^2}^{(0)}&=1-128\delta\Delta_{\phi^2}^{(2)} \label{eqn: constraint 2 from the O(epsilon) analysis} \ .
\end{align}
These constraints relate the higher-order bCFT data of $\phi^2$ and $\partial^2\phi^2$. Consequently, our boundary bootstrap results indicate operator mixing between $\partial^2\phi^2$ and the level-two descendant $\Box\phi^2$ of $\phi^2$ in the interacting theory. We will investigate this operator mixing using conventional perturbation theory in Section~\ref{subsec: A Puzzle with partial^2phi^2}.

\subsection{A Puzzle with $\partial^2\phi^2$}
\label{subsec: A Puzzle with partial^2phi^2}

In this section, we clarify some possible confusions associated with the primary $\partial^2\phi^2$ in our $d=6-\epsilon$ interacting $\Box^2$ CFT. In particular, we rederive its peculiar $O(\epsilon^{-1})$ anomalous OPE coefficient $\delta\lambda a_{\partial^2\phi^2}^{(-1)}$ using conventional perturbation theory. More generally, we discuss some features that are exclusive to the $\epsilon$ expansion around the dimension where the staggered module arises.

\begin{figure}[ht]
    \centering
    \begin{tikzpicture}[scale=3]
        \draw[->] (0, 0) -- (2, 0) node[right] {$d$};
        \draw[->] (0, 0) -- (0, 2) node[above] {$g$};
        \draw[blue] (1.5, 1.5) -- (0, 1.5) node[left, black] {$0$};
        \draw[red, domain=0:1.5, samples=100] plot (\x, {0.444*(\x*\x)+0.5});
        \draw[dotted] (1.5, 1.5) -- (1.5, 0) node[below] {$d_0$};
        \draw[dotted] (0.75, 1.5) -- (0.75, 0) node[below] {$d_0-\epsilon$};
        \draw[dotted] (0.75, 0.75) -- (0, 0.75) node[left] {$g_\star(\epsilon)$};
        \draw [fill=black] (1.5,1.5) circle[radius= 0.06 em] node[above right] {$S^{(0)}$};
        \draw [fill=blue] (0.75,1.5) circle[radius= 0.06 em] node[above, blue] {$S_\text{free}^{(\epsilon)}$};
        \draw [fill=red] (0.75,0.75) circle[radius= 0.06 em] node[below right, red] {$S_\text{WF}^{(\epsilon)}$};
    \end{tikzpicture}
    \caption{Two lines of CFTs ended on a $d_0$ dimensional free CFT $S^{(0)}$. The blue line is a line of free CFTs, while the red curve (shape indicated qualitatively) is a line of interacting CFTs. 
    }
    \label{fig: Relation between three CFTs.}
\end{figure}
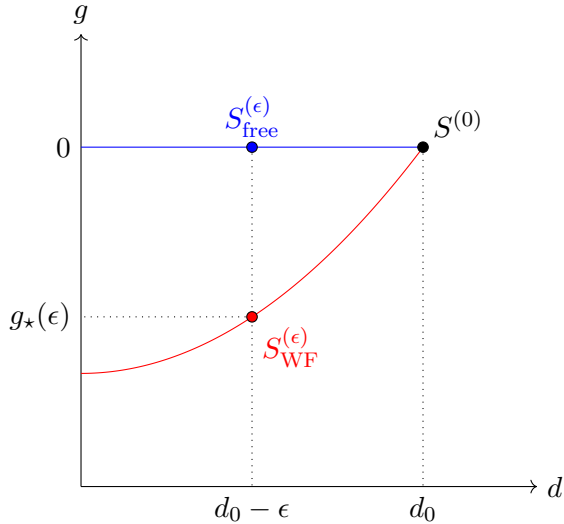

As illustrated in figure~\ref{fig: Relation between three CFTs.}, $S^{(0)}$ is a free CFT in $d_0$ dimensions, and $S^{(\epsilon)}$ are CFTs in $d_0-\epsilon$ dimensions obtained by smoothly deforming $S^{(0)}$. We adopt \textbf{Axiom II} from \cite{Rychkov:2015naa} as the definition of a smooth deformation: \textit{Correlators of operators $[\mathcal{O}_i]$ in $S^{(\epsilon)}$ approach correlators of operators $\mathcal{O}_i$ in $S^{(0)}$ in the $\epsilon\to 0$ limit:} 
\begin{equation}
\label{eqn: axiom 2 of rychkov and tan}
    \lim_{\epsilon\to 0}\braket{[\mathcal{O}_1](x_1)[\mathcal{O}_2](x_2)\cdots}=\braket{\mathcal{O}_1(x_1)\mathcal{O}_2(x_2)\cdots} \ .
\end{equation}
There is always a trivial smooth deformation $S_\text{free}^{(\epsilon)}$: the free CFT in $d_0-\epsilon$ dimensions. We are interested in the non-trivial one, $S_\text{WF}^{(\epsilon)}$, where the subscript WF is used because the simplest example is the Wilson-Fisher fixed point. $S_\text{WF}^{(\epsilon)}$ can be distinguished from $S_\text{free}^{(\epsilon)}$ by specifying the equation of motion.

It seems to be an immediate corollary that the $S^{(\epsilon)}$ CFT data should approach the $S^{(0)}$ CFT data in the $\epsilon \to 0$ limit. However, $\lambda a_{\partial^2\phi^2}$ provides a counterexample. Taking $S^{(0)}$ to be the $d=6$ free $\Box^2$ CFT, $\lambda a_{\partial^2\phi^2}$ is not defined in $S^{(0)}$. In the $d=6-\epsilon$ free theory $S_\text{free}^{(\epsilon)}$, $\lambda a_{\partial^2\phi^2} = \lambda a_{\partial^2\phi^2}^{(0)} \sim \epsilon^{-1}$ is well-defined, but its $\epsilon \to 0$ limit does not exist. One might wonder whether $S_\text{free}^{(\epsilon)}$ is still a smooth deformation of $S^{(0)}$. The answer is yes because these properties of $\lambda a_{\partial^2\phi^2}$ are consistent with the smoothness condition \eqref{eqn: axiom 2 of rychkov and tan}. More explicitly, recall that the OPE coefficient should be CFT-normalized
\begin{equation} 
\label{eqn: dependence of lambda a on norm} 
\lambda a_{\partial^2\phi^2} \propto \frac{1}{\braket{\partial^2\phi^2|\partial^2\phi^2}} \ .
\end{equation}
Since $\partial^2\phi^2=\frac{1}{2}\Box\phi^2$ is a primary descendant in $S^{(0)}$, it has zero norm, and thus $\lambda a_{\partial^2\phi^2}$ is not defined. The norm \eqref{eqn: norm of d^2 phi^2} of $\partial^2\phi^2$ in $S_\text{free}^{(\epsilon)}$ is of order $\epsilon$, so $\lambda a_{\partial^2\phi^2} \sim \epsilon^{-1}$ is well-defined. This norm vanishes as $\epsilon \to 0$, satisfying the smoothness condition for the two-point correlator of $\partial^2\phi^2$.

Assume our $d=6-\epsilon$ interacting CFT is a smooth deformation of $S^{(0)}$ and denote it by $S^{(\epsilon)}_\text{WF}$. The smoothness condition alone requires only that the norm of $\partial^2\phi^2$ in $S^{(\epsilon)}_\text{WF}$ vanishes as $\epsilon \to 0$, but its asymptotic behavior does not need to match that of the norm in $S_\text{free}^{(\epsilon)}$. In other words,
\begin{equation}
\label{eqn: ratio of correlators in free and WF does not have to be 1}
    \lim_{\epsilon\to 0}\frac{\braket{[\partial^2\phi^2](x)[\partial^2\phi^2](y)}}{\braket{\partial^2\phi^2(x)\partial^2\phi^2(y)}_0}=c
\end{equation}
where the numerator is the correlator in $S^{(\epsilon)}_\text{WF}$, the denominator is the correlator in $S^{(\epsilon)}_\text{free}$, and $c$ is a constant that does not have to be one. In contrast, if $\mathcal{O}$ were a primary whose norm is $\mathcal{N}_\mathcal{O}\neq 0$ in $S^{(0)}$, then its norm must be of the form $\mathcal{N}_\mathcal{O} + O(\epsilon)$ in both smooth deformations $S^{(\epsilon)}_\text{free}$ and $S^{(\epsilon)}_\text{WF}$. As a result,
\begin{equation}
\label{eqn: ratio of normal correlators has to be 1}
    \lim_{\epsilon\to 0}\frac{\braket{[\mathcal{O}](x)[\mathcal{O}](y)}}{\braket{\mathcal{O}(x)\mathcal{O}(y)}_0}=1 \ .
\end{equation}
From \eqref{eqn: dependence of lambda a on norm}, we note that
\begin{equation}
\label{eqn: ratio of lambda a does not have to be 1}
    c=\lim_{\epsilon\to 0}\frac{\lambda a_{\partial^2\phi^2}\text{ in }S^{(\epsilon)}_\text{free}}{\lambda a_{\partial^2\phi^2}\text{ in }S^{(\epsilon)}_\text{WF}}=\lim_{\epsilon\to 0}\frac{\lambda a_{\partial^2\phi^2}^{(0)}}{\lambda a_{\partial^2\phi^2}^{(0)}+\delta \lambda a_{\partial^2\phi^2}^{(-1)}\epsilon^{-1}}
    \ .
\end{equation}
Hence, there will be an $O(\epsilon^{-1})$ anomalous OPE coefficient of $\partial^2\phi^2$ if $c\neq 1$. We will rederive \eqref{eqn: ratio of lambda a for partial^2 phi^2}, or equivalently, $c = 1 - 2\delta\Delta_{\phi^2}^{(1)}$, by carefully renormalizing $\partial^2\phi^2$ in $S^{(\epsilon)}_\text{WF}$.

From the perspective of conventional perturbation theory, it is not immediately obvious that the renormalization of $\partial^2\phi^2$ in $S^{(\epsilon)}_\text{WF}$ requires more care than, say, the renormalization of $\phi^2$. To motivate our results, we first review renormalization in the minimal subtraction scheme. The well-known formula \eqref{eqn: CPT formula for anomalous dimension} produces a wrong anomalous dimension of $\partial^2\phi^2$. Moreover, a naive application of this procedure to $\Box\phi^2$ leads to a paradox --- as a descendant of $\phi^2$, its anomalous dimension seems to differ from that of $\phi^2$. Resolving this paradox will provide insights into how to renormalize $\partial^2\phi^2$, since $\partial^2\phi^2$ differs from $\frac{1}{2}\Box\phi^2$ only by an $O(\epsilon)$ term.

All correlators $\braket{\mathcal{O}_1(x_1)\mathcal{O}_2(x_2)\cdots}_0$ in $S^{(\epsilon)}_\text{free}$ are known. We want to compute the renormalized correlators $\braket{[\mathcal{O}_1](x_1)[\mathcal{O}_2](x_2)\cdots}$ in 
\begin{equation}
    S^{(\epsilon)}_\text{WF}=S^{(\epsilon)}_\text{free}+\int\diff^dx\,g_\star\mu^\epsilon\mathcal{O}_\text{int}(x)
\end{equation}
where $\mathcal{O}_\text{int}$ is a weakly relevant primary in $S^{(\epsilon)}_\text{free}$ with scaling dimension $\Delta_1=d-\epsilon$. For our purposes, we focus on the two-point correlator of a scalar primary $\mathcal{O}$ with scaling dimension $\Delta$. The bare correlator in $S^{(\epsilon)}_\text{WF}$ is
\begin{equation}
    \braket{\mathcal{O}(x)\mathcal{O}(y)}=\braket{\mathcal{O}(x)\mathcal{O}(y)}_0-g_0\int\diff^dz\,\braket{\mathcal{O}(x)\mathcal{O}(y)\mathcal{O}_\text{int}(z)}_0+O(g_0^2)
\end{equation}
where $g_0=\mu^\epsilon\left(g+O(g^2)\right)$ is the bare coupling and
\begin{equation}
    \braket{\mathcal{O}(x)\mathcal{O}(y)}_0=\frac{\mathcal{N}_\mathcal{O}}{(x-y)^{2\Delta}},\quad \braket{\mathcal{O}(x)\mathcal{O}(y)\mathcal{O}_\text{int}(z)}_0=\frac{C_{\mathcal{O}\mathcal{O}\mathcal{O}_\text{int}}}{(x-y)^{2\Delta-\Delta_1}(y-z)^{\Delta_1}(x-z)^{\Delta_1}} \ .
\end{equation}
The one-loop integral can be evaluated using the formula
\begin{equation}
    \int\diff^dz\,\frac{1}{(x-z)^{2\alpha}(y-z)^{2\beta}}=\frac{S_d}{(x-y)^{2(\alpha+\beta-d/2)}}\frac{\Gamma(\frac{d}{2})\Gamma(\frac{d}{2}-\alpha)\Gamma(\frac{d}{2}-\beta)\Gamma(\alpha+\beta-\frac{d}{2})}{2\Gamma(\alpha)\Gamma(\beta)\Gamma(d-\alpha-\beta)}
\end{equation}
to obtain
\begin{equation}
    \braket{\mathcal{O}(x)\mathcal{O}(y)}=\braket{\mathcal{O}(x)\mathcal{O}(y)}_0\biggl[1-g S_d\frac{C_{\mathcal{O}\mathcal{O}\mathcal{O}_\text{int}}}{\mathcal{N}_\mathcal{O}}\left(\frac{2}{\epsilon}+\log[\mu^2(x-y)^2]+O(\epsilon)\right)+O(g^2)\biggl] \ .
\end{equation}
Assuming there is no operator mixing, we define the renormalized operator $[\mathcal{O}]=Z^{-1}\mathcal{O}$ by absorbing the $\epsilon$ poles into the wavefunction renormalization of $\mathcal{O}$:
\begin{equation}
\label{eqn: CPT formula for wavefunction renormalization}
    Z=1-\frac{g S_d C_{\mathcal{O}\mathcal{O}\mathcal{O}_\text{int}}}{\mathcal{N}_\mathcal{O}\epsilon} +O(g^2) \ .
\end{equation}
The anomalous dimension 
\begin{equation}
\label{eqn: CPT formula for anomalous dimension}
    \gamma=\frac{g S_d C_{\mathcal{O}\mathcal{O}\mathcal{O}_\text{int}}}{\mathcal{N}_\mathcal{O}}  +O(\epsilon^2)
\end{equation}
of $\mathcal{O}$ can be read off by comparing the renormalized correlator
\begin{equation}
    \left.\braket{[\mathcal{O}](x)[\mathcal{O}](y)}\right\rvert_{g=g_\star}=\braket{\mathcal{O}(x)\mathcal{O}(y)}_0\left(1-g_\star S_d\frac{C_{\mathcal{O}\mathcal{O}\mathcal{O}_\text{int}}}{\mathcal{N}_\mathcal{O}} \log(x-y)^2+\cdots\right)
\end{equation}
with the standard form of the CFT two-point correlator. Or equivalently, $\gamma$ is related to $Z$ by a standard RG argument:
\begin{equation}
\label{eqn: standard formula for anomalous dimension}
    \gamma=\left.\pdv{\log Z}{\log \mu}\right\rvert_{g=g_\star} \ .
\end{equation}
As a straightforward application of \eqref{eqn: CPT formula for anomalous dimension}, we can recover $\delta\Delta_{\phi^2}^{(1)}=3/8$ after accounting for the $\kappa$ factor \eqref{eqn: kappa factor}, which allows us to work with the CFT-normalized $\phi$:
\begin{equation}
\label{eq:limitvals}
    g_\star=\kappa\left(-\frac{3\epsilon}{8}+O(\epsilon^2)\right),\quad S_d=\frac{2\pi^\frac{d}{2}}{\Gamma(d/2)},\quad \frac{C_{\phi^2\phi^2\mathcal{O}_\text{int}}}{\mathcal{N}_{\phi^2}}=-8(d-4),\quad d=6-\epsilon \ .
\end{equation}
However, applying \eqref{eqn: CPT formula for anomalous dimension} to $\partial^2\phi^2$ gives $\delta\Delta_{\partial^2\phi^2}^{(1)}=0$ because
\begin{equation}
    \braket{\partial^2\phi^2(x)\partial^2\phi^2(y)\mathcal{O}_\text{int}(z)}_0=0 \ .
\end{equation}
This contradicts our bootstrap result $\delta\Delta_{\partial^2\phi^2}^{(1)}=3/8$. To understand why $\delta\Delta_{\partial^2\phi^2}^{(1)}=0$ is incorrect, it is instructive to first consider the renormalization of $\Box\phi^2$.

Since $\Box\phi^2$ is not a primary, we cannot use \eqref{eqn: CPT formula for wavefunction renormalization} and \eqref{eqn: CPT formula for anomalous dimension}. Therefore, we take a step back and compute the bare correlator
\begin{equation}
    \begin{split}
        \braket{\Box\phi^2(x)\Box\phi^2(y)}&=\braket{\Box\phi^2(x)\Box\phi^2(y)}_0\biggl[1\\
        &+gS_d\frac{24(d-5)(3d-14)}{d-3}\left(\frac{2}{\epsilon}+\log[\mu^2(x-y)^2]+O(\epsilon)\right)+O(g^2)\biggl]
        \ . 
    \end{split}
\end{equation}
Naively, we would choose
\begin{equation}
    Z_{\Box\phi^2}=1+\frac{32gS_d}{\epsilon}+O(g^2)
\end{equation}
so that the $2/\epsilon$ term is removed from the renormalized correlator
\begin{equation}
    \left.\braket{[\Box\phi^2](x)[\Box\phi^2](y)}\right\rvert_{g=g_\star}=\braket{[\Box\phi^2](x)[\Box\phi^2](y)}_0\left(1-\frac{3\epsilon}{4}\log(x-y)^2+\cdots\right)
\ .
\end{equation} 
The $O(\epsilon)$ anomalous dimension of $\Box\phi^2$ can then be read off, or computed using \eqref{eqn: standard formula for anomalous dimension}, to be
\begin{equation}
    \delta\Delta_{\Box\phi^2}^{(1)}=\frac{3}{4}\neq \frac{3}{8}=\delta\Delta_{\phi^2}^{(1)} \ .
\end{equation}
This is paradoxical because $\Box\phi^2$, as a level-2 descendant of $\phi^2$, must have
\begin{equation}
    \Delta_{\Box\phi^2}=\Delta_{\phi^2}+2\quad \Leftrightarrow\quad \delta\Delta_{\Box\phi^2}^{(n)}=\delta\Delta_{\phi^2}^{(n)}\quad\text{for all }n\geq 1 \ .
\end{equation}
The resolution to this paradox can be found in our earlier discussion, where we argued that if an operator has zero norm in $S^{(0)}$, then the constant $c$ in \eqref{eqn: ratio of correlators in free and WF does not have to be 1} does not have to be one. In the $d=6$ free $\Box^2$ CFT, the norm of $\Box\phi^2=2\partial^2\phi^2$ is zero. Thus, we can have $c\neq1$ in
\begin{equation}
    \left.\braket{[\Box\phi^2](x)[\Box\phi^2](y)}\right\rvert_{g=g_\star}=\braket{\Box\phi^2(x)\Box\phi^2(y)}_0\left(c+O(\epsilon)\right) \ .
\end{equation}
We determine $c$ to be
\begin{equation}
    c=1-2\delta\Delta_{\phi^2}^{(1)}=\frac{1}{4}
\end{equation}
by first acting $\Box_x\Box_y$ on the renormalized $\phi^2$ correlator
\begin{equation}
\label{eqn: renormalized Box phi^2 correlator to O(epsilon)}
    \left.\braket{[\Box\phi^2](x)[\Box\phi^2](y)}\right\rvert_{g=g_\star}=\Box_x\Box_y\left(\frac{\mathcal{N}_{\phi^2}}{(x-y)^{2\Delta_{\phi^2}}}\right)
\end{equation}
and then expanding around $\epsilon=0$. This suggests that we should choose
\begin{equation}
\label{eqn: wavefunction renormalization of Box phi^2}
    Z_{\Box\phi^2}=1+\frac{16gS_d}{\epsilon}+O(g^2)
\end{equation}
so that 
\begin{equation}
\label{eqn: renormalized Box phi^2 correlator to O(g)}
    \left.\braket{[\Box\phi^2](x)[\Box\phi^2](y)}\right\rvert_{g=g_\star}=\braket{[\Box\phi^2](x)[\Box\phi^2](y)}_0\left(\frac{1}{4}-\frac{3\epsilon}{4}\log(x-y)^2+\cdots\right)
    \ .
\end{equation}
Substituting \eqref{eqn: wavefunction renormalization of Box phi^2} into \eqref{eqn: standard formula for anomalous dimension}, we obtain $\delta\Delta_{\Box\phi^2}^{(1)}=3/8=\delta\Delta_{\phi^2}^{(1)}$ as desired. However, note that we can no longer read off $\delta\Delta_{\Box\phi^2}^{(1)}$ by comparing our $O(g)$ calculation \eqref{eqn: renormalized Box phi^2 correlator to O(g)} with the expansion of \eqref{eqn: renormalized Box phi^2 correlator to O(epsilon)} around $\epsilon=0$. This is because all terms of the form
\begin{equation}
    \frac{g^{n+1}}{\epsilon^{n}}\log(x-y)^2\subset \braket{[\Box\phi^2](x)[\Box\phi^2](y)}
\end{equation}
could contribute to the coefficient of $\epsilon\log(x-y)^2$, and therefore, higher-loop calculations are required. We have not carried out this exercise explicitly, but it would be interesting to do.

We will now renormalize $\partial^2\phi^2$ in $S^{(\epsilon)}_\text{WF}$. Our starting point is the wavefunction renormalizations:
\begin{equation}
    Z_{\phi^2}=1+\frac{16gS_d}{\epsilon}+O(g^2),\quad Z_{\partial^2\phi^2}=1+\frac{16gS_d}{\epsilon}+O(g^2)
\end{equation}
which recover the bootstrap results $\delta\Delta_{\phi^2}^{(1)}=\delta\Delta_{\partial^2\phi^2}^{(1)}=3/8$. We first show that there is finite mixing between the bare primaries $\phi^2$ and $\partial^2\phi^2$. As suggested in \cite{Berenstein:2016avf}, we can modify $\partial^2\phi^2$ with a finite counterterm proportional to $\Box\phi^2$ to remove the mixing. More precisely, $\partial^2\phi^2$ should be renormalized as follows:
\begin{equation}
\label{eqn: renormalized d^2 phi^2}
    [\partial^2\phi^2]=Z_{\partial^2\phi^2}^{-1}\left(\frac{1}{2}\Box\phi^2+\left(1-2\delta\Delta_{\phi^2}^{(1)}\right)\frac{d-6}{2}\phi\Box\phi\right)
\end{equation}
in order to maintain the orthogonality between the renormalized primaries $[\phi^2]$ and $[\partial^2\phi^2]$ to $O(g)$. Now, the constant $c$ in \eqref{eqn: ratio of correlators in free and WF does not have to be 1} can be straightforwardly computed to be 
\begin{equation}
    c=1-2\delta\Delta_{\phi^2}^{(1)}
\end{equation}
recovering the bootstrap result \eqref{eqn: ratio of lambda a for partial^2 phi^2}.

We demonstrate finite mixing between $\phi^2$ and $\partial^2\phi^2$ by contradiction. Suppose there is no mixing. In that case, the renormalized primaries are $[\phi^2] = Z_{\phi^2}^{-1} \phi^2$ and $[\partial^2\phi^2] = Z_{\partial^2\phi^2}^{-1} \partial^2\phi^2$. These operators fail to be orthogonal because, to $O(g)$, the renormalized correlator
\begin{equation}
    \braket{[\phi^2](x)[\partial^2\phi^2](y)}=\frac{4d(3d-14)(d-4)gS_d}{(x-y)^{2(d-3)}}\left(-\frac{2}{3}+O(\epsilon)\right)+O(g^2)
\end{equation}
is non-zero. Since the renormalization $[\phi^2] = Z_{\phi^2}^{-1} \phi^2$ leads to the correct renormalized correlator
\begin{equation}
    \left.\braket{[\phi^2](x)[\phi^2](y)}\right\rvert_{g=g_\star}=\frac{2}{(x-y)^{2(d-4)}}\left(1-\frac{3\epsilon}{8}\log(x-y)^2+\cdots\right)
\end{equation}
and anomalous dimension $\delta\Delta_{\phi^2}^{(1)} = 3/8$, we must modify the renormalization of $\partial^2\phi^2$ to eliminate the mixing. We propose the ansatz
\begin{equation}
    [\partial^2\phi^2]=\frac{1}{1+2\alpha}Z_{\partial^2\phi^2}^{-1}\left(\partial^2\phi^2+\alpha \Box\phi^2\right)
\end{equation}
where the normalization factor $(1 + 2\alpha)^{-1}$ ensures that the bare operator $Z_{\partial^2\phi^2} [\partial^2\phi^2]$ approaches $\frac{1}{2} \Box \phi^2$ in the $\epsilon \to 0$ limit. Now, the renormalized correlator becomes
\begin{equation}
\label{eqn: renormalized correlator of phi^2 and d^2 phi^2}
\begin{split}
    \braket{[\phi^2](x)[\partial^2\phi^2](y)}=&\frac{1}{1+2\alpha}\left(1+\frac{16gS_d}{\epsilon}+O(g^2)\right)^{-2}\frac{1}{(x-y)^{2(d-3)}}\biggl[4(d-4)(d-6)\alpha\\
    &+4d(3d-14)(d-4)gS_d\left(-\frac{2}{3}+O(\epsilon)\right)\\
    &+32(3d-14)(d-4)(d-6)\alpha gS_d\left(\frac{2}{\epsilon}+O(\epsilon^0)\right)+O(g^2)\biggl] \ .
\end{split}
\end{equation}
In the square bracket, the first term is $\braket{\phi^2(x)\Box\phi^2(y)}_0$, the second term is the $O(g)$ correction to $\braket{\phi^2(x)\partial^2\phi^2(y)}$, and the third term is the $O(g)$ correction to $\braket{\phi^2(x)\Box\phi^2(y)}$. Keeping terms to $O(g)$ and evaluating at 
(\ref{eq:limitvals}), we find that $[\phi^2]$ is orthogonal to $[\partial^2\phi^2]$ only if
\begin{equation}
    \alpha=\frac{3}{2}=\frac{\delta\Delta_{\phi^2}^{(1)}}{1-2\delta\Delta_{\phi^2}^{(1)}}\quad\Leftrightarrow\quad[\partial^2\phi^2]=(1-2\delta\Delta_{\phi^2}^{(1)})Z_{\partial^2\phi^2}^{-1}\left(\partial^2\phi^2+\frac{\delta\Delta_{\phi^2}^{(1)}}{1-2\delta\Delta_{\phi^2}^{(1)}}\Box\phi^2\right)
\end{equation}
which is equivalent to \eqref{eqn: renormalized d^2 phi^2}.

The constant $c$ in \eqref{eqn: ratio of correlators in free and WF does not have to be 1} can be determined by calculating the renormalized correlator
\begin{equation}
    \begin{split}
        \braket{[\partial^2\phi^2](x)[\partial^2\phi^2](y)}=&\frac{1}{(1+2\alpha)^2}\left(1+\frac{16gS_d}{\epsilon}+O(g^2)\right)^{-2}\braket{\partial^2\phi^2(x)\partial^2\phi^2(y)}_0\\
        &\times \biggl[1-\frac{8(d-3)\alpha^2}{d}-16(3d-14)(d-5)\alpha gS_d\left(\frac{2}{\epsilon}+O(\epsilon^0)\right)\\
        &-\frac{192(3d-14)(d-5)\alpha^2 gS_d}{d}\left(\frac{2}{\epsilon}+O(\epsilon^0)\right)+O(g^2)\biggl] \ .
    \end{split}
\end{equation}
In the square bracket, the first term is $\braket{\partial^2\phi^2(x)\partial^2\phi^2(y)}_0$, the second term is $\braket{\Box\phi^2(x)\Box\phi^2(y)}_0$, the third term is the $O(g)$ correction to $\braket{\partial^2\phi^2(x)\Box\phi^2(y)}$, and the fourth term is the $O(g)$ correction to $\braket{\Box\phi^2(x)\Box\phi^2(y)}$. Keeping terms to $O(g)$ and evaluating at
(\ref{eq:limitvals}) and $\alpha = 3/2$, we find
\begin{equation}
    \left.\braket{[\partial^2\phi^2](x)[\partial^2\phi^2](y)}\right\rvert_{g=g_\star}=\braket{\partial^2\phi^2(x)\partial^2\phi^2(y)}_0\left(\frac{1}{4}+O(\epsilon)\right) \ .
\end{equation}
In other words,
\begin{equation}
    c=\frac{1}{4}=1-2\delta\Delta_{\phi^2}^{(1)}
    \ .
\end{equation}

\section{Discussion}
\label{sec:discussion}

We have commenced here an analysis of higher derivative scalar conformal 
field theories with interactions and boundary.
Looking at a simple example (\ref{action}) with a $\Box^2$ kinetic term and quartic coupling, 
we observed some relations between one-loop anomalous dimensions.  In the bulk, 
the relation
\[
\delta \Delta_{\phi^2}^{(1)} = \delta \Delta_{\partial^2 \phi^2}^{(1)} 
\]
can be traced to the existence of a staggered module in the $d=6$ theory.  The equality comes from the constraint
that the associated conformal blocks must combine together in a non-singular way in $d=6$.  Indeed, we found a whole host of relations between the $\phi^2$ and $\partial^2 \phi^2$ conformal data, (\ref{eqn: ratio of lambda a for partial^2 phi^2}),  
(\ref{eqn: O(epsilon) anomalous dimensions relation}), 
(\ref{eqn: constraint 1 from the O(epsilon) analysis}), 
and (\ref{eqn: constraint 2 from the O(epsilon) analysis}), associated with the existence of this staggered module.

The symmetry in the boundary anomalous dimensions in table 
\ref{tab: Boundary anomalous dimensions for the four conformal boundary conditions.} 
on the other hand came from a combination of the facts that these boundary operators are shadow dual at tree level, the anomalous dimensions are linearly proportional to coefficients in the free Green's function, and these coefficients change sign in a simple way when moving from $NN$ to $DD$ or from $ND$ to $DN$ boundary conditions.  A similar relation exists (for a similar reason) between 
$\delta \hat \Delta_{\phi}^{(1)} = \delta \hat \Delta_{\partial_n \phi}^{(1)}$ 
in $\phi^4$ theory in $4-\epsilon$ 
dimensions which does not persist at next order in the epsilon expansion \cite{reeve1981renormalisation,reeve1980critical,diehl1981field,Bissi:2018mcq}.  

There is much we have not addressed.  For example, we could add a $\phi^6$ interaction in the bulk and consider the two additional fixed points that are accessible because of its addition \cite{Safari:2017tgs,Safari:2017irw}, but  now in the presence of a boundary.

More interesting perhaps is the role of boundary conditions.
The emphasis in the work has been on Dirichlet and Neumann type boundary conditions.  However, in the presence of interactions, a third type of boundary condition becomes available, associated with so-called
extraordinary critical behavior.  
Classically, one can see from solving the equations of motion that follow from the Lagrangian density in $d=6$
\[
{\mathcal L} = \frac{1}{2} \phi \Box^2 \phi + 
\frac{g}{2} \phi^2 \Box \phi^2
+h \phi^6
\]
that $\phi = \frac{c}{x_n}$ is a solution provided 
\[
4 + 2 c^2 g + c^4 h = 0 \ .
\]
For the case we studied, where $h=0$, $c = \sqrt{-2/g}$ which is real provided $g<0$.  It would be interesting to try to extend the work here to this surface ordered case in the future.

A way to access this extraordinary type of boundary condition and possibly other conformal boundary conditions as well is through the addition of boundary interactions.  
Similar to what happens when a $\phi^2$ boundary term is added to the $\phi^4$ theory in $4-\epsilon$ dimensions \cite{Diehl:1996kd},
ref.\ \cite{Chalabi:2022qit} argued that in the free theory, the relevant boundary mass terms
$(\Phi^{(0)})^2$ and $(\Phi^{(1)})^2$ will change a Neumann to a Dirichlet boundary condition.  For example, $(\Phi^{(0)})^2$ in the $ND$ theory will produce a flow to the $DD$ theory.  Interactions should not change the story, at least in perturbation theory.
Moreover, it is known that adding a $\phi^2$ boundary term with the opposite sign in $\phi^4$ theory leads to extraordinary behavior \cite{Diehl:1996kd}.  In analogy, adding $(\Phi^{(0)})^2$ or $(\Phi^{(1)})^2$ with the wrong sign may also lead to extraordinary behavior  in this higher derivative setting.

Another class of boundary interactions considered in \cite{Chalabi:2022qit} promotes the boundary to an interface between two theories with different boundary conditions on each side.  A classically marginal interaction 
of the form $\Phi^{(i)}_R \Phi^{(3-i)}_L$ then acts to rotate the boundary conditions away from Neumann or Dirichlet on each side.  It may be interesting to consider how interactions affect this story.  
Based on what happens with the anomalous dimensions of the boundary operators $\phi$ and $\partial_n \phi$ 
in $\phi^4$ theory in $4-\epsilon$ dimensions, it seems likely that this boundary interaction will fail to be exactly marginal starting at $O(\epsilon^2)$.  We can ask if it will lead to new real fixed points.

So far, the boundary interactions discussed are all quadratic.  Two final proposals for boundary interactions are the classically marginal $\phi^5$ and $\phi^3 \partial_n \phi$.  They are expected to lead to short RG flows, precisely because they are classically marginal.  Their effect could be studied perturbatively in an $\epsilon$ expansion, perhaps leading to new real fixed points.

It is also interesting to study higher derivative theories in the presence of a defect with codimension greater than one. While performing a defect bootstrap is difficult --- due to the lack of a closed-form expression for the bulk conformal block in the most general case \cite{Billo:2016cpy} --- a conventional perturbative approach suffices to compute the $\epsilon$ expansions of defect CFT data. For example, we can consider the $d=6-\epsilon$ $\Box^2$ CFT, both in its free and interacting forms, in the presence of a surface defect with a $\phi^2$ defect term. This setup serves as a higher-derivative analogue of the models analyzed in \cite{Trepanier:2023tvb, Giombi:2023dqs, Raviv-Moshe:2023yvq}. To further investigate the $\epsilon$ expansion around the dimension where the staggered module arises, we could also consider a codimension-two defect hosting the peculiar primary $\partial^2\phi^2$.

To end with some cosmological remarks, we make a connection with the appearance of late-time logarithms in de Sitter space (see for example \cite{Anninos:2014lwa,Cohen:2020php,Gorbenko:2019rza}).  
Qualitatively, we should be able to Wick rotate our framework, turning the normal coordinate $x_n$ into time, and then through a Weyl rescaling turn our space-time into de Sitter space, with the far future corresponding to the limit $t \to 0$ from below.
Thinking of time then as the normal direction $x_n$, we saw a very similar type of near boundary logarithm in our perturbative discussion of the $\langle \phi \phi \rangle$ two-point function.  The coefficients of these logarithms yield corrections to anomalous dimensions of boundary operators $\Phi^{(i)}$.  Resumming the perturbative expansion should ultimately give power law scaling, given by the actual resummed scaling dimensions of these boundary operators.  It is reassuring to see this structure persists regardless of whether the underlying bCFT is unitary or not.  It would be interesting to explore these logarithms and other parallels with QFT in de Sitter space in the future.

\section*{Acknowledgments}
We would like to thank Dio Anninos, Shivaji Sondhi, Andy Stergiou, and
Arkady Tseytlin for discussion.
C.H.\ thanks the Oxford University Physics Department for hospitality, where much of this work was prepared.
This work was partially supported by a Wolfson Fellowship from the Royal Society and by the
U.K. Science and Technology Facilities Council Grant ST/P000258/1.

\appendix

\section{Some Cross-Checks}\label{app: Cross-Check from Conventional Perturbation Theory}

In this appendix, we verify the critical coupling $g_\star = - 3 \epsilon /8 + O(\epsilon^2)$ (\ref{eqn: critical coupling}) and the vanishing of the OPE coefficient
 $\delta \lambda a_6^{(1)} = 0$ \eqref{eqn: NN result of the O(epsilon) OPE of 6}.
We also show that $\delta\Delta_{\partial^2\phi^2}^{(1)}=3/8$ recovers the $O(\epsilon^2)$ anomalous dimension of $\phi$ computed in \cite{Safari:2017irw, Safari:2017tgs}.

\subsection*{Critical Coupling}

For completeness, and to avoid any confusion caused by the different normalization conventions used in our work and in \cite{Safari:2017irw, Safari:2017tgs}, here we present an independent derivation of (\ref{eqn: critical coupling}). The one-loop beta function for the coupling $g$ in \eqref{eqn: action again} is given by the standard result of the conformal perturbation theory \cite{Cardy:1996xt}
\begin{equation}
\label{eqn: beta function}
    \beta_g=-\epsilon g+\frac{1}{2}S_{d}\frac{C_3}{C_2}g^2+O(g^3)
\end{equation}
where 
\begin{equation}
C_3=\braket{\mathcal{O}_\text{int}\mathcal{O}_\text{int}\mathcal{O}_\text{int}}_0=64(19d^3-174d^2+416d^2-80)\left[\frac{1}{16\pi^\frac{d}{2}}\Gamma\left(\frac{d-4}{2}\right)\right]^{6}
\end{equation} 
and
\begin{equation}
    C_2=\braket{\mathcal{O}_\text{int}\mathcal{O}_\text{int}}_0=-8 d (3 d - 14)\left[\frac{1}{16\pi^\frac{d}{2}}\Gamma\left(\frac{d-4}{2}\right)\right]^{4}
\end{equation}
are the three-point and two-point correlators in the free CFT $S_0$ (omitting the standard position-dependent factors), computed using the canonically normalized scalar field $\phi_\text{can}$
\begin{equation}
    \braket{\phi_\text{can}(x)\phi_\text{can}(y)}_0=\frac{1}{16\pi^\frac{d}{2}}\Gamma\left(\frac{d-4}{2}\right)\frac{1}{(x-y)^{2\Delta_\phi}} \ .
\end{equation}
Evaluating \eqref{eqn: beta function} at $d=6-\epsilon$, the non-trivial root of the one-loop beta function is
\begin{equation}
    g_\star=-6\pi^3\epsilon+O(\epsilon^2)\quad \text{(canonical normalization)}  \ .
\end{equation}
In this work, we assume the scalar field $\phi$ is CFT-normalized
\begin{equation}
    \braket{\phi(x)\phi(y)}_0=\frac{1}{(x-y)^{2\Delta_\phi}} \ .
\end{equation}
Thus, the $\epsilon$ expansion of $g_\star$ that we will use is
\begin{equation}
    g_\star=-\frac{3\epsilon}{8}+O(\epsilon^2)\quad \text{(CFT normalization)}  \ .
\end{equation}

\subsection*{That $\delta \lambda a_6^{(1)} = 0$}

Recall that 
\begin{equation}
\label{eqn: 2 terms in the O(epsilon) OPE of 6}
    \delta\lambda a_6^{(1)}=\delta \lambda a_{\phi^6}^{(1)}+\delta \lambda a_{\mathcal{O}_\text{int}}^{(1)} \ .
\end{equation}
Neither $\phi^6$ nor $\mathcal{O}_\text{int}$ appear in the free $\phi\times\phi$ OPE, so
\begin{equation}
    \lambda_{\phi^6}^{(0)}=\lambda_{\mathcal{O}_\text{int}}^{(0)}=0\quad\Rightarrow\quad \delta\lambda a_{\phi^6}^{(1)}=\delta\lambda_{\phi^6}^{(1)}a_{\phi^6}^{(0)},\quad \delta\lambda a_{\mathcal{O}_\text{int}}^{(1)}=\delta\lambda_{\mathcal{O}_\text{int}}^{(1)}a_{\mathcal{O}_\text{int}}^{(0)} \ .
\end{equation}
Since the bulk OPE is a local property of the bulk CFT and therefore unaffected by the presence of a boundary, the $O(\epsilon)$ OPE coefficient $\delta\lambda_{\phi^6}^{(1)}$ can be extracted from the $O(g_\star)$ correction to the three-point correlator $\braket{\phi\phi\phi^6}$ in the interacting $\Box^2$ CFT without a boundary
\begin{equation}
    \braket{\phi(x)\phi(y)\phi^6(z)}=-g_\star\int\diff^du\,\braket{\phi(x)\phi(y)\phi^6(z)\mathcal{O}_\text{int}(u)}_0+O(g_\star^2) \ .
\end{equation}
We can verify that $\braket{\phi\phi\phi^6\mathcal{O}_\text{int}}_0=0$, implying $\delta\lambda_{\phi^6}^{(1)}=0$. 
 At the level of the free theory, $\braket{\phi\phi\phi^6\mathcal{O}_\text{int}}_0 = 30\langle \phi \phi \rangle_0^2 \langle \phi^4 {\mathcal O}_{\rm int} \rangle_0$ and $\langle \phi^4 {\mathcal O}_{\rm int} \rangle_0$ vanishes by orthogonality because both $\phi^4$ and ${\mathcal O}_{\rm int}$ are primary operators in the free theory.
The $O(\epsilon^0)$ one-point coefficient $a_{\mathcal{O}_\text{int}}^{(0)}$ can be extracted from the free one-point function $\braket{\mathcal{O}_\text{int}}_0$ in $d=6$. Under the $NN$ boundary condition,
\begin{equation}
    \braket{\mathcal{O}_\text{int}(x)}_0=\left.-\frac{d(d-2)(d-6)}{(2x_n)^{2d-6}}\right\rvert_{d=6}=0 \ .
\end{equation}
Therefore, $a_{\mathcal{O}_\text{int}}^{(0)}=0$ and both terms in \eqref{eqn: 2 terms in the O(epsilon) OPE of 6} are zero, verifying \eqref{eqn: NN result of the O(epsilon) OPE of 6}.

\subsection*{That $\delta\Delta_{\partial^2\phi^2}^{(1)}=3/8$}

We use the method developed in \cite{Rychkov:2015naa} to provide a cross-check of $\delta\Delta_{\partial^2\phi^2}^{(1)}=3/8$. Consider the interacting $\partial^2\phi^2\times \phi$ OPE. 
From \eqref{Cijk}, we see that terms
with $m>1$ and $n=0$ all have a $1/\epsilon$ pole.  We focus on the leading divergent term with $m=1$ and $n=0$:
\begin{equation}
    \partial^2\phi^2(x)\times \phi(0)\supset \lambda_{\partial^2\phi^2}x^{-\Delta_{\partial^2\phi^2}}\left[1+\cdots+\left( -\frac{\delta\Delta_{\partial^2\phi^2}^{(1)}}{256\delta\Delta_\phi^{(2)}\epsilon}+O(\epsilon^0)\right)x^4\Box^2+\cdots\right]\phi(0) \ .
\end{equation}
Using the interacting equation of motion
\eqref{eqn: interacting EOM}, we replace $\Box^2 \phi$ to yield
\begin{eqnarray}
   \lefteqn{ \partial^2\phi^2(x)\times \phi(0)\supset } \\
   &&
   \lambda_{\partial^2\phi^2}x^{-\Delta_{\partial^2\phi^2}}\left[\phi(0)+\cdots+\left( -\frac{3\delta\Delta_{\partial^2\phi^2}^{(1)}}{1024\delta\Delta_\phi^{(2)}}+O(\epsilon)\right)x^4\left(\phi^2\Box\phi+\frac{1}{3}\Box\phi^3\right)(0)+\cdots\right] \ . \nonumber
\end{eqnarray}
In the $\epsilon\to 0$ limit,
\begin{equation}
\label{eqn: interacting (d phi)^2 phi OPE in the epsilon to 0 limit}
    \partial^2\phi^2(x)\times\phi(0)\supset -4 x^{-4}\left[\phi(0)+\cdots -\frac{3\delta\Delta_{\partial^2\phi^2}^{(1)}}{1024\delta\Delta_\phi^{(2)}}x^4\left(\phi^2\Box\phi+\frac{1}{3}\Box\phi^3\right)(0)+\cdots\right]
\end{equation}
we expect \eqref{eqn: interacting (d phi)^2 phi OPE in the epsilon to 0 limit} to become the $d=6$ free $\partial^2\phi^2\times\phi$ OPE. 
Recall from \eqref{eqn: bulk primary phi (d phi)^2} that $\phi^2\Box\phi+\frac{1}{3}\Box\phi^3$ is a $d=6$ primary.
(The higher order $1/\epsilon$ terms
with $m>1$ that we neglected above are descendants
of this primary in the OPE.)
We contract both sides of this OPE with
$\phi^2\Box\phi+\frac{1}{3}\Box\phi^3$ in the free theory to find a constraint on the OPE coefficient:
\begin{equation}
1=\frac{3\delta\Delta_{\partial^2\phi^2}^{(1)}}{128\delta\Delta_\phi^{(2)}} \ .
\end{equation}
Substituting $\delta\Delta_{\partial^2\phi^2}^{(1)}=3/8$, we obtain
\begin{equation}
    \delta\Delta_\phi^{(2)}=\frac{9}{1024}
\end{equation}
which agrees with the computation in \cite{Safari:2017irw, Safari:2017tgs}, providing a posteriori justification of our $\partial^2\phi^2$ $O(\epsilon)$ anomalous dimension.

\newpage
\bibliographystyle{JHEP}
\bibliography{bib}

\end{document}